\documentclass[prd,letterpaper, amsmath, amssymb, preprintnumbers, showpacs, floatfix, nofootinbib, superscriptaddress]{revtex4-1}
\usepackage{tikz}
\usepackage{mystyle_emacs}
\usepackage{stackrel}
\usepackage{theoremref}
\usepackage{placeins}
\usepackage{multirow}
\usepackage{graphicx}
\usepackage{xcolor}\usepackage[labelformat=empty]{subcaption}
\newcommand{\psibar}{\bar \psi}

\def\bp{{\bb \phi}}
\def\vbp{{\vec{\bb \phi}}}
\def\vJ{{\vec J}}
\def\vp{{\vec \phi}}

\def\oRp{{\omega_{R,+}}}
\def\oRn{{\omega_{R,-}}}
\def\oI{{\omega_I}}
\def\xr{3}
\def\yr{3}
\usetikzlibrary{decorations.markings,decorations.pathmorphing}

\newcommand\newsubcap[1]{\phantomcaption%
       \caption*{\textbf{#1}}}
\newcommand\capa{\textbf{(a)}\,}
\newcommand\capb{\textbf{(b)}\,}
\newcommand\capc{\textbf{(c)}\,}

\def\op{{\bullet}}
\begin{document}

\title{Correlation function distributions for O(N) lattice field theories in the disordered phase}
\author{Cagin Yunus}
\email{cyunus@mit.edu}
\affiliation{Center for Theoretical Physics,
Massachusetts Institute of Technology, Cambridge, MA 02139, USA}
\author{William Detmold}
\email{wdetmold@mit.edu}
\affiliation{Center for Theoretical Physics,
Massachusetts Institute of Technology, Cambridge, MA 02139, USA}
\affiliation{The NSF Institute for Artificial Intelligence and Fundamental Interactions}

\begin{abstract}
Numerical computations in strongly-interacting quantum field theories are often performed using Monte-Carlo sampling methods. A key task in these calculations is to estimate the value of a given physical quantity from the distribution of stochastic samples that are generated using the Monte-Carlo method. Typically, the sample mean and sample variance are used to define the expectation values and uncertainties of computed quantities. However, the Monte-Carlo sample distribution contains more information than these basic properties and it is useful to investigate it more generally.
In this work, the exact form of the probability distributions of two-point correlation functions at zero momentum in $O(N)$ lattice field theories in the disordered phase and in infinite volume are determined. These distributions allow for a robust investigation of the efficacy of the Monte-Carlo sampling procedure and are shown also to allow for improved estimators of the target physical quantity to be constructed. The theoretical expectations are shown to agree with numerical calculations in the $O(2)$ model.

\end{abstract}

\preprint{{MIT-CTP-5550}}
\maketitle
%\tableofcontents

%\newpage
\section{Introduction}

Quantum field theories (QFTs) are used pervasively in theoretical particle, nuclear, and condensed-matter physics to describe a wide range of physical phenomena. 
In many cases, these QFTs are strongly interacting and numerical methods are needed in order to make predictions. In many cases, the only effective numerical approach is to discretise and compactify spacetime and use importance-sampling Monte Carlo to estimate  the Euclidean path integrals that result. In many situations, this is a stochastically challenging task; numerical results for correlation functions between fields at different points usually exhibit  signal-to-noise ratios that degrade exponentially in the separation between the points. This makes extraction of physical information such as the masses of excitations of the theory difficult. To confront these difficulties, various strategies for the reduction of variance in the Monte Carlo procedure have been pursued \cite{DellaMorte:2007zz,DellaMorte:2008jd,DellaMorte:2010yp,Detmold:2014hla,Majumdar:2014cqa,Ce:2016ajy,Ce:2016idq,Wagman:2017xfh,Wagman:2017gqi,Detmold:2018eqd,Porter:2016vry,DallaBrida:2020cik,Detmold:2020ncp,Kanwar:2021wzm,Detmold:2021ulb}. Previous works have also investigated the nature of statistical fluctuations in specific QFTs empirically \cite{Beane:2009gs,Endres:2011jm,Endres:2011mm,DeGrand:2012ik,Grabowska:2012ik,Nicholson:2012xt,Drut:2015uua,Wagman:2016bam}.  
In Ref.~\cite{Yunus:2022pto}, the first analytic insights into the structure of the probability distribution functions (PDFs) of correlation functions have been presented in the context of a scalar field theory.  
In general, knowledge of the probability distributions can lead to more efficient estimators of the mean.\footnote{For example, for a uniform distribution on the interval $[0,a]$ for $a>0$, the minimum-variance unbiased estimator of the mean, $\hat \m_{U,mvue} = \ff{\caln+1}{2\caln} \max_i x_i$, 
can be shown to outperform the sample mean by using specific properties of the PDF.} In the context of Monte Carlo sampling, knowledge of the PDF can also allow for statistical tests of thermalisation and auto-correlation of the samples. 

In this work, the PDFs of two-point correlation functions in the lattice $O(N)$ model in the disordered phase at vanishing spatial momentum will be calculated, significantly extending the results in Ref.~\cite{Yunus:2022pto}. 
$O(N)$ models are a generalization of the Ising model and have been studied extensively since their initial definition in Ref. \cite{inception}. These models exhibit many important phenomena; for example in two dimensions, the $O(2)$ model features a Berezinskii–Kosterlitz–Thouless transition at low temperatures, while $O(N)$ models for $N>2$ are asymptotically free and are therefore strongly coupled at long distances, see \cite{kogut} for a review. In any dimension, the $O(N)$ model reduces to a self-avoiding random walk as $N\to0$ as first demonstrated in \cite{DEGENNES1972339}. The $O(N)$ model therefore provides an interesting testing ground for extensions of the previous work on PDFs of correlation functions and forms the central basis of this work. 

The structure of this work is as follows.
In Sec. \ref{sec:Free}, the probability distributions of two point correlation functions at vanishing spatial momentum are calculated for a free complex scalar field, equivalent to the $O(2)$ model. In Sec. \ref{Sec:ON}, this is extended to the case of interacting $O(N)$ models in the disordered phase and the derived form is shown to hold at all temporal separations of the correlation function. In Sec. \ref{sec:numerical},  numerical results are presented for interacting $O(2)$ models in two dimensions in the disordered phase. These results confirm the predictions of Sec. \ref{Sec:ON} and provide support for the assumptions needed to derive the distributions. 
Using the PDFs of two-point correlation functions in the $O(2)$ model, an improved estimator of the mean is constructed in Sec.\ref{sec:improved_estimator}. Sec. \ref{sec:conc} summarizes the results and presents an outlook.

\section{Free Complex Field}
\label{sec:Free}

To introduce the calculation of correlation function probability distributions, the case of a single free complex scalar field is first discussed, generalizing the discussion of the real scalar field in Ref.~\cite{Yunus:2022pto}.  
Let $\psi(t,\vec x) = \phi_1(t,\vec x) + i \phi_2(t,\vec x)$ be a free  complex field, where $\phi_i(t,\vec x)\in \mathbb{R}$. A $d+1$  dimensional free Euclidean lattice field theory will be considered with $L$ being the size of each of the spatial directions in lattice units and $\b$ being the size of the temporal direction. For a free lattice field theory, volume averaged\footnote{In what follows, fields with bars will denote spatially averaged fields: $\bar \co(t) = L^{-\ff d 2} \sum_{\vec x}\co(t,\vec x)$. The scaling with $L^{-\ff d 2}$ is fixed by requiring that $\bar \co(t)$ has finite and non-zero variance as $L\to\ii$.} fields $\bb \phi_{i}(t)$ will decouple from other modes due to  momentum conservation. Therefore, the partition function for this sector of the theory can be written as:
\bad 
Z = \int \cald \bb \phi_1 e^{-\ff 1 2 \bb \phi_1^T D \bb \phi_1} \int \cald \bb \phi_2 e^{-\ff 1 2 \bb \phi_2^T D \bb \phi_2} 
\ead  
where the vector notation 
\be 
\bb \phi_{1,2} = \bpm \bb \phi_{1,2}(0) \\ \cds \\ \bb \phi_{1,2}(\b-1) \epm
\ee 
is used, with $\cald \bp_{i} = \prod_{t'=0}^{\b-1}d\bp_{i}(t')$. $D$ is a real symmetric matrix arising from the kinetic operator after integrating over the non-zero momentum modes; the exact form of $D$ is not relevant in what follows.  Since all integrals are Gaussian, one may choose to integrate out all of the field variables except $\bb \phi_1(0)$, $\bb \phi_1(t)$, $\bb \phi_2(0)$ and $\bb \phi_2(t)$ (in what follows, only the temporal separation between the fields is important, so, the temporal location of one pair of fields is chosen to be at 0 without loss of generality).   The partition function  can then be expressed as: 
\bad 
Z =& \caln(t) \int d\bb \phi_1(0) d\bb \phi_1(t) d\bb \phi_2(0) d\bb \phi_2(t)   \exp{-\ff{\s_f^2}{2} \ff{\lp \bb \phi_1(0)^2 + \bb \phi_1(t)^2 + \bb \phi_2(0)^2 + \bb \phi_2(t)^2\rp}{ \s_f^4 - K_f(t)^2 } }\\ 
&\hspace*{6cm} \times \exp{\ff{K_f(t)}{\s_f^4-K_f(t)^2}\lp \bb \phi_1(0) \bb \phi_1(t) + \bb \phi_2(0) \bb \phi_2(t)\rp},
\label{eq:Z_cartesian}
\ead
 where
 \bad 
 \s_f^2 &= \ev{\bar \phi_1(0)\bar \phi_1(0)}
 = \ev{\bar \phi_2(0)\bar \phi_2(0)},\\
 K_f(t) &= \ev{\bar \phi_{1}(t)\bar \phi_{1}(0)} 
 =\ev{\bar \phi_{2}(t)\bar \phi_{2}(0)},
 \label{eq:sKdef}
 \ead
and the normalization factor $\caln(t)$ can be calculated by performing the remaining Gaussian integrals exactly and imposing that $Z=1$: 
\be
\caln(t) = \ff{1}{(2\pi)^2(\s_f^4-K_f(t)^2)}.
\ee 
For $Z$ defined by Eq. \eqref{eq:Z_cartesian} to be finite, the condition $K_f(t)<\s_f^{-2}$ must be satisfied. Furthermore, the inequality $K_f(t)>0$ follows from the reflection positivity and Eq. \eqref{eq:sKdef}. 

It follows from this partition function that the joint probability distribution of $\bb \phi_1(0), \bb \phi_1(t), \bb \phi_2(0), \bb \phi_2(t)$ taking values $u_0, u_t, v_0, v_t$ respectively is given by
\begin{widetext}
\bad 
P_{\bb \phi_1(0),\bb \phi_1(t),\bb \phi_2(0), \bb \phi_2(t)}(u_0,u_t,v_0,v_t) &=\ff{1}{(2\pi)^2(\s_f^4-K_f(t)^2)} \\ &\quad \times\exp{-\ff{\s_f^2}{2\lp \s_f^4 - K_f(t)^2 \rp}\lp  u_0^2 + u_t^2 + v_0^2 +  v_t^2\rp + \ff{K_f(t)}{\s_f^4-K_f(t)^2}\lp u_0 u_t + v_0 v_t \rp}.
\label{eq:free_joint_pdf}
\ead 
\end{widetext}
Since $\ev{\bp_{i}(t)\bp_{i}(0)}\propto e^{-mt}$ for $i\in\{1,2\}$ and $t \gg \ff{1}{\ti m - m}$, where $m$ is the energy of the first excited state and $\ti m$ is the energy of the second excited state with vanishing spatial momentum, it follows that $K_f(t) \propto e^{-mt}$ in the same large time limit. The distributions of 
\be 
C_{Re}(t) = Re(\psibar(t)\psibar^*(0)) = \bp_1(t)\bp_1(0)+\bp_2(t)\bp_2(0)
\ee 
and 
\be  
C_{Im}(t) = Im(\psibar(t)\psibar^*(0))= \bp_2(t)\bp_1(0)-\bp_1(t)\bp_2(0)
\ee
can be calculated from this joint probability density. The corresponding characteristic functions\footnote{The characteristic function $\Phi_{A}(\o)$ of a random variable $A$ is defined as $\Phi_A(\o) = \int d x e^{-i\o x}P_A(x)$, where $P_A(x)$ is the probability density function of $A$.} can be calculated from Eq. \eqref{eq:free_joint_pdf}, giving:

\bad 
\Phi_{C_{Re}(t)}(\o) &= \expval{e^{-i\o C_{Re}(t) }} \\ 
&= \expval{e^{-i\o \bp_1(t)\bp_1(0)}}\expval{e^{-i\o\bp_2(t)\bp_2(0)}} \\ 
&= \ff{\oRp(t) \oRn(t)}{\lp \o - i\oRp(t) \rp \lp \o + i \oRn(t) \rp},\\ 
\Phi_{C_{Im}(t)}(\o) &= \ff{\oI(t)^2}{\lp \o - i \oI(t) \rp \lp \o+ i\oI(t) \rp},
\ead 
where the independence of $\bp_1(t)$ and $\bp_2(t)$ has been used and  $\oRp(t), \oRn(t), \oI(t)$ are defined by:
\bad 
\o_{R,+}(t) &= \ff{1}{\s_f^2+K_f(t)}, \\ 
\o_{R,-}(t) &= \ff{1}{\s_f^2-K_f(t)}, \\ 
\o_I(t) &= \ff{1}{\sqrt{\s_f^4-K_f(t)^2}}.
\ead 

Consequently,the probability distributions $P_{C_{Re,Im}(t)}(x) = \int d\o  e^{i\o x} \ti P_{C_{Re,Im}(t)}(\o)$  are given by:
\bad 
P_{C_{Re}(t)}(x) &= \begin{cases}
\ff{\oRp(t) \oRn(t)}{\oRp(t)+\oRn(t)}e^{-\oRp(t) x} \quad &\text{ for } x > 0 \\ 
\ff{\oRp(t) \oRn(t)}{\oRp(t)+\oRn(t)}e^{\oRn(t) x} \quad  &\text{ for } x<0 
\end{cases}, \\ 
P_{C_{Im}(t)}(x) &= \ff{\oI(t)}{2}e^{-\oI(t) \abs{x}}.
\label{eq:C_2d_dists}
\ead 

Since there is strong evidence that phase fluctuations rather than magnitude fluctuations are central to signal-to-noise degradation in many QFT contexts \cite{Wagman:2017xfh,Wagman:2017gqi,Detmold:2018eqd}, 
it is also of interest to calculate $P_{R_t}(x)$ and $P_{\theta_t}(x)$ where $R_t = \abs{\bb \psi(t)\bb \psi^*(0)}$ is the magnitude and $\theta_t = \arg\lp \bb \psi(t) \bb \psi^*(0)\rp$ is the phase of the correlation function. For this purpose, it is useful to express the partition function in Eq. \eqref{eq:Z_cartesian} in terms of the polar fields defined by the relations
\bad 
r_0 &= \ss{\phi_1(0)^2+\phi_2(0)^2}, \\ 
\xi_0 &= \arctan{\ff{\phi_2(0)}{\phi_1(0)}}, \\ 
r_t &= \ss{\phi_1(t)^2 +\phi_2(t)^2}, \\ 
\xi_t & = \arctan{\ff{\phi_2(t)}{\phi_1(t)}}.
\ead 
With this substitution, Eq. \eqref{eq:Z_cartesian} becomes:
\bad 
Z &= \ff{1}{(2\pi)^2(\s_f^4-K_f(t)^2)}\int dr_0 dr_t d\xi_0 d\xi_t r_0 r_t \\ 
&\qquad \qquad \qquad \times \exp{-\ff{\s_f^2}{2\lp \s_f^4 - K_f(t)^2 \rp} \lk \lp r_0^2 + r_t^2 \rp + 2\ff{K_f(t)}{\s_f^2}r_0r_t\cos{\xi_t-\xi_0} \rk}.
\ead
Since this is independent of $\xi_0+\xi_t$,  the partition function can be integrated over this combination to give
\bad 
Z &= \ff{1}{2\pi(\s_f^4-K_f(t)^2)}\int dr_0 dr_t d\t_t r_0 r_t \exp{-\ff{\s_f^2}{2\lp \s_f^4 - K_f(t)^2 \rp} \lk \lp r_0^2 + r_t^2 \rp + 2\ff{K_f(t)}{\s_f^2}r_0r_t\cos{\t_t} \rk}, 
\label{eq:Z_polar}
\ead 
where  $\t_t = \xi_t - \xi_0$.

To find $P_{\t_t}(x)$, one integrates over $r_0$ and $r_t$ in Eq. \eqref{eq:Z_polar} to obtain:
\bad 
Z &= \ff{1-\ff{K_f(t)^2}{\s_f^4}}{4\pi} \int \ff{d\t_t}{ \lp 1-\ff{K_f(t)^2}{\s_f^4}\cos{\t_t}^2\rp^2} \Bigg[ 2 \lp 1-\ff{K_f(t)^2}{\s_f^4}\cos{\t_t}^2\rp  +\pi \ff{K_f(t)}{\s_f^2} \ss{1-\ff{K_f(t)^2}{\s_f^4}\cos{\t_t}^2}\cos{\t_t} \\ &\qquad \qquad  +2\cos{\t_t}\ff{K_f(t)}{\s_f^2}\ss{1-\ff{K_f(t)^2}{\s_f^4}\cos{\t_t}^2}+\arctan{\ff{K_f(t)^2\cos{\t_t}}{\s_f^4\ss{1-\ff{K_f(t)}{\s_f^2}\cos{\t_t}^2}}} \Bigg]\,.
\label{eq:phase_func}
\ead 
For $K_f(t) \ll \s_f^{-2}$, Eq. \eqref{eq:phase_func} can be expressed as a series in ${K_f(t)}/{\s_f^2}$ as
\bad 
Z = \int d\t_t \lp \ff{1}{2\pi} + \ff{K_f(t)}{4\s_f^2}\cos{\t_t} + \co\lp \lp \ff{K_f(t)}{\s_f^2}\rp^2 \rp \rp
\ead 
and therefore, the angular probability density is
\bad 
P_{\t_t}(x) = 
    \ff{1}{2\pi}+ \ff{K_f(t)}{4\s_f^2}\cos{x} + \co \lp \lp \ff{K_f(t)}{\s_f^2}\rp^2\rp,
\ead 
where the domain of $x$ is $-\pi \leq x < \pi$. 

To find $P_{R_t}(x)$, one integrates over $\t_t$ in Eq. \eqref{eq:Z_polar} and can define $R_t = r_0 r_t$, leading to
\bad 
Z &= \ff{1}{\s_f^4-K_f(t)^2}\int dr_0 dr_t r_0 r_t I_0(K_f(t)r_0r_t)e^{-\ff{1}{2\s_f^2} \lp r_0^2 + r_t^2 \rp } \\ 
&=\ff{1}{\s_f^4-K_f(t)^2}\int dR_t dr_0 dr_t r_0 r_t \d\lp R_t - r_0 r_t \rp I_0(K_f(t)r_0r_t)e^{-\ff{1}{2\s_f^2} \lp r_0^2 + r_t^2 \rp } \\
&= \ff{1}{\s_f^4-K_f(t)^2}\int_0^\ii dR_t\, I_0 \lp K_f(t)R_t \rp K_0\lp \ff{R_t}{\s_f^2} \rp R_t,
\ead 
after $r_{0,t}$ are integrated out, where $I_n(x)$ and $K_n(x)$ are modified Bessel functions of the first and second kind, respectively. Therefore, the radial probability density is
\bad 
P_{R_t}(x) = \ff{1}{\s_f^4-K_f(t)^2} I_0 \lp K_f(t)x \rp K_0\lp \ff{x}{\s_f^2} \rp x.
\ead

\section{Probobility Distributions of $O(N)$ models in the disordered phase}
\label{Sec:ON}

In this section, an interacting Euclidean lattice field theory with $N$ real bosonic fields, $\phi_{a}(t,\vec x)$ where $a \in \{1,\cds,N\}$ will be considered such that
\begin{itemize}
\item The Euclidean action of the theory is real,
\item There is a unique, translationally invariant, gapped vacuum $\ket{\O}$,
\item $\phi_a(t,\vec x)$ is covariant under temporal and spatial translations for $a \in \{1,\cds, N\}$,
\item The vacuum expectation value of $\phi_a(t,\vec x)$, $\expval{\phi_a(t,\vec x)}{\O}$, vanishes for $a \in \{1,\cds,N\}$.
\end{itemize}

In the following subsections, probability distributions of the volume averaged fields defined at different times will be presented\footnote{Some of the arguments used in this section are closely related to the proofs of the Central Limit Theorem for random vectors, see for example Ref. \cite{Van_der_Vaart2000-fz}.} in increasing generality. While the methods work in the general case, some of the results will be presented for the specific case where the theory is invariant under an $O(N)$ symmetry relating the $N$ fields.
As in the previous section, $L$ is the extent of each of the $d$ spatial directions and $\beta$ is the extent of the temporal direction.

\subsection{Large-time Limit}\label{sec:LargeTime}

In this subsection, it will be shown that the distributions of two-point correlation functions in $O(N)$ models in the disordered phase have a universal large-time limit.

% In Ref.~\cite{Yunus:2022pto}, the case $N=1$ was considered. Denoting the $N=1$ field by $\phi(t,\vec x)$, it was shown that
% \bad 
% \lim_{L \to \ii} \mel{\O}{\d \lp \bp(0)-u\rp}{\O} = \ff{1}{\ss{2\pi}\s_\phi}e^{-\ff{u^2}{2\s_\phi^2} },
% \label{eq:ex_paper_formula}
% \ead 
% for some $\s_\phi>0$. It was also found that the PDF of the two-point correlation function $C(t)=\bp(t)\bp(0)$ is given by
% \bad
% \lim_{t \to \ii} \lim_{\b \to \ii} \lim_{L\to \ii} P_{\bp(t)\bp(0)}(x) = \ff{1}{\pi \s^2_\phi}K_0\lp \ff{\abs{x}}{\s^2_\phi}\rp.
% \label{eq:prod_dist_N=1}
%  \ead 

%Eq. \eqref{eq:ex_paper_formula} can be used  to generalize the findings of Ref.~\cite{Yunus:2022pto} to $N>1$. 

The components of the (spatially-averaged) $N$-component field can be collected as a vector 
%Define $\bb \phi_a(t)$ and $\vbp (t)$ by
\bad 
%\bb \phi_a(t) &= L^{-\ff d 2}\sum_{\vec x}\phi_a(t,\vec x), \\ 
\vbp(t) &=  \bpm \bp_{1}(t) \\ \vdots \\ \bp_{N}(t)\epm 
\ead 
and a natural object to consider is the PDF $P_{\vec q \cd \vbp(0)}(u)$ that represent the probability $\vec q \cd \vbp(0)$ takes the value $u$ for some fixed vector $\vec q \in \mathbb{R}^N$. In the limit where the temporal extent $\b$ is taken to infinity, this is given by 
\bad 
P_{\vec q \cd \vbp(0)}(u) &= \lim_{\b \to \ii} \Tr\lp e^{-\b H} \d \lp \vec q \cd \vbp(0)-u \rp \rp \\ 
&= \expval{\d \lp \vec q \cd \vbp(0)-u \rp}{\O}.
\ead 

In Ref.~\cite{Yunus:2022pto}, the large-time behaviour of the analogous quantity for a single field was derived. Generalising this to the present case leads to:
\bad 
\lim_{L \to \ii} \mel{\O}{\d \lp \vec q \cd \vbp(0)-u\rp}{\O} = \ff{1}{\ss{2\pi}\s(\vec q)}e^{-\ff{u^2}{2\s(\vec q)^2} }.
\label{eq:qphi}
\ead 
Here, the quantity $\sigma(\vec q)$ can be shown to be given by:
% \bad 
% \s(\vec q)^2 &= \ff{1}{\ss{2\pi}\s(\vec q)}\int du\, u^2e^{-\ff{u^2}{2\s(\vec q)^2} } \\ 
% &= \lim_{L \to \ii} \int du\, u^2 \mel{\O}{\d \lp \vec q \cd \vbp(0)-u\rp}{\O}\\
% &= \lim_{L\to \ii} \int du\, u^2 \int \cald \vp\, \abs{\bk{\vp}{\O}}^2 \d \lp \vec q \cd \vbp(0)-u\rp \\
% &= \lim_{L\to\ii} \int \cald \vp\, \abs{\bk{\vp}{\O}}^2 \lp \vec q \cd \vp \rp^2 \\
% &=\lim_{L\to \ii} \mel{\O}{\lp\vec q \cd \vbp(0) \rp \lp \vec q \cd \vbp(0) \rp }{\O} %\\ 
% %&= \sum_{a,b=1}^N q_a q_b \lim_{L\to\ii}\mel{\O}{\bb \phi_{a}(0)\bb \phi_{b}(0)}{\O}.
% \\ 
% &=  \vec q^T \Sigma \vec q
% \ead
\bad 
\sigma(\vec q)^2 &=\vec q^T \Sigma \vec q
\ead
in terms of a symmetric matrix $\Sigma$ with components
\be 
\Sigma_{ab} =  \lim_{L\to\infty}\mel{\O}{\bb \phi_{a}(0)\bb \phi_{b}(0)}{\O}.
\ee
%where $\cald \vp = \prod_{\vec x}d\phi(\vec x,0).$

The characteristic function 
\be  
\Phi_{\vec q \cd \vbp(0)}(\o) = \int dx e^{-i\omega x}  P_{\vec q \cd \vbp(0)}(x)
\ee   
of $\vec q \cd \vbp(0)$
evaluated at $\o=1$ can be calculated  in the large volume limit as:
\bad 
\Phi_{\vec q \cd \vbp(0)}(1) &=  \lim_{L \to \ii} \mel{\O}{e^{-i \vec q \cd \vbp(0)}}{\O} \\ 
&= e^{-\ff 1 2 \s(\vec q)^2} \\
&= e^{-\ff 1 2  \vec q^T \Sigma \vec q},
\ead 
where the second equality follows from Eq.~\eqref{eq:qphi}.

By making use of the numerical equivalence between   the characteristic function of $\vbp$ evaluated at $\vec q$ and the characteristic function of $\vec q \cd \vbp(0)$ evaluated at $1$, this leads to:
\bad 
\Phi_{\vbp(0)}(\vec q) &= \lim_{L \to \ii} \mel{\O}{e^{-i\vec q \cd \vbp}}{\O}\\ 
&= \Phi_{\vec q \cd \vbp(0)}(1)\\
&= e^{-\ff 1 2 \vec q^T \Sigma \vec q}.
\ead 

By performing the inverse Fourier transform the large volume probability density function $P_{\vbp(0)}(\vec u)$ of $\vbp(0)$ can be determined and is given by

\bad 
P_{\vbp(0)}(\vec u) = \ff{1}{\ss{(2\pi)^N\det \Sigma}}e^{-\ff 1 2 {\vec u}^T \inv \Sigma \vec u }.
\label{eq:Pu}
\ead 
Thus the probability distribution of an $N$-component field is an $N$-dimensional correlated Gaussian. This will be useful in defining more complicated joint- and product-distributions.

The joint probability distribution of $\vbp(t)$ and $\vbp(0)$ taking vector values $\vec u$ and $\vec v$ respectively is defined by
\bad
P_{\vbp(t),\vbp(0)}(\vec u,\vec v)&=   \lim_{\b \to \ii} Tr\left[ e^{-\b H}\d \lp \vbp(t)-\vec u \rp  \d \lp \vbp(0)-\vec v \rp \right] \\
&=\sum_n e^{-E_n t} \mel{\O}{\d \lp \vbp(0) - \v u \rp}{n} \mel{n}{\d \lp \vbp(0)-\v v \rp}{\O}  \\
&\equiv \sum_n P^{(n)}_{\vbp(t),\vbp(0)}(\vec u,\vec v)
\label{eq:Pjoint},
\ead
where each term in the summation in the last line arises from the contributions of successively higher energy states. That is, $E_n\leq E_{n+1}$ with $E_0=0$ and the sum is over states $|n\rangle$ with vanishing spatial momentum such that $\mel{\O}{\d \lp \vbp(t)-\vec u\rp}{n} \neq 0$ for some $\vec u \in \mathbb{R}^N$. 

In the large time limit, only the vacuum intermediate state is relevant in Eq.~\eqref{eq:Pjoint} and 
the leading term is given by
\bad 
P^{(0)}_{\vbp(t),\vbp(0)}(\vec u,\vec v) = \lim_{t\to\infty} P_{\vbp(t),\vbp(0)}(\vec u,\vec v) 
\\ 
&= \mel{\O}{\d\lp \vbp(t)-\vec u\rp}{\O}\mel{\O}{\d \lp \vbp(0)-\vec v\rp}{\O}   \\ 
&= P_{\vbp(0)}(\vec u)P_{\vbp(0)}(\vec v) \\
&= \ff{1}{(2\pi)^N \det \S}e^{-\ff 1 2  \vec u^T \inv \S \vec u - \ff 1 2 \vec v^T \inv \S \vec v}.
\label{eq:PSigmauv}
\ead

In the case of a theory with $O(N)$ symmetry, further simplifications occur as it can be shown\footnote{The fact that $\Sigma_{ab}$ is diagonal follows  from invariance under the $O(N)$ transformation  $\phi_a(t,\vec x) \to -\phi_a(t,\vec x)$ for fixed $a$ with $\phi_b(t,\vec x)$ invariant for $b \neq a$. The fact that $\S$ is proportional to the identity follows from the requirement of invariance under the $O(N)$ transformation $\phi_a(t,\vec x) \to -\phi_b(t,\vec x)$, $\phi_b(t,\vec x) \to \phi_a(t,\vec x)$ for fixed $a$ and $b$ with  $\phi_c(t,\vec x)$ invariant for $c \notin \{a,b\}$.} for that case that 
 \bad 
 \Sigma_{ab} = \s^2 \d_{ab},
 \label{eq:sigma_O(N)_def}
 \ead 
 for some $\s>0$.

 For $N>2$, the only $O(N)$-invariant quadratic bilocal operator is $C_{\op}(t) = \sum_{a=1}^N \bb \phi_a(t)\bb\phi_a(0)$ and the product distribution for this quantity is
\bad
P_{C_{\op}(t)}(x) \equiv \int d\vec u d\vec v 
P_{\vbp(t),\vbp(0)}(\vec u,\vec v) \delta(\vec u\cdot \vec v-x),
\ead
which, as in Eq.~\eqref{eq:Pjoint}, can be expanded in terms of the contributions of intermediate states as
\bad 
P_{C_{\op}(t)}(x) =\sum_n P^{(n)}_{C_{\op}(t)}(x) .
\label{eq:PCdot}
\ead
The large time limit of $P_{C_\op}(t)$ will be derived below, see  Eq. \eqref{eq:asym_func}. 

Similarly, the characteristic function of $C_{\op}(t)$ can be expanded in partial contributions as:
\bad 
\Phi_{C_{\op}(t)}(\o) &\equiv 
\int dx\, e^{-i\o x}P_{C_{\op}(t)}(x)\\
&= \sum_n\Phi^{(n)}_{C_{\op}(t)}(\o)(\vec u,\vec v).
\ead

\subsection{Leading Corrections}
\label{sec:leading}

In this section, the leading correction $P^{(1)}_{\vbp(t),\vbp(0)}(\v u,\v v)$ in $P_{\vbp(t),\vbp(0)}(\v u,\v v)$ away from the $t \to \ii$ limit will be derived. For this purpose, it is assumed that the first multiplet of excited states with vanishing spatial momentum transform in the fundamental representation of $O(N)$ and may be labelled as $\ket {m,a}$ for $a \in \{1,\cds,N\}$. The state $\ket {m,a}$ can be created by acting\footnote{The factor of $e^{-mT}$ is included so that in an expansion of $e^{-mT}\bp_a(T)\ket \O$ in terms of the eigenstates of the theory, the coefficient $\ket{m,a}$ is independent of $T$ and the coefficients of all other terms vanish in the limit $T \to -\ii$.} with $\bp_a(t)$ on the vacuum

\be
\ket{m,a} = \lim_{T \to -\ii} e^{-mT} \bp_a(T)\ket \O
\label{eq:m_a_bp}
\ee
as $\mel{\O}{\bp_a(0)}{\O} = 0$ by assumption. The leading correction as $t \to \ii$ is given by 
\bwt
\bad 
P^{(1)}_{\vbp(t),\vbp(0)}(\v u, \v v) &= e^{-mt}\sum_a \mel{\O}{\d \lp \vbp(0) - \v u \rp}{m,a} \mel{m,a}{\d \lp \vbp(0)-\v v \rp}{\O}.  
\ead 
\\

To calculate this expression, $\mel{\O}{\d \lp \vbp(0) - \v u \rp}{m,a}$ must be determined. From Eq. \eqref{eq:m_a_bp}, it follows that
\bad 
\mel{\O}{\d \lp \vbp(0)- \v u \rp}{m,a} = \lim_{T \to -\ii} e^{-mT}\mel{\O}{\d \lp \vbp(0) - \v u \rp \bp_a(T)}{\O}.
\label{eq:m_to_phi}
\ead 
The insertion of $\bp_a(T)$ can be obtained through the response to a time-dependent, but spatially constant source term as:
\bad
\mel{\O}{\d \lp \vbp(0) - \v u \rp \bp_a(T)}{\O} &= \ff{\pp}{\pp J_a}\mel{\O}{\d \lp \vbp(0) - \v u \rp e^{\vJ \cd \vbp(T)}}{\O}\Big\rvert_{\vJ = 0}.
\label{eq:phi_to_J}
\ead
For infinite temporal extent, $\mel{\O}{\d \lp \vbp(0) - \v u \rp e^{\vJ \cd \vbp(T)}}{\O}$ is equal to a path integral expression:

\bad 
\mel{\O}{\d \lp \vbp(0) - \v u \rp e^{\vJ \cd \vbp(T)}}{\O} = \lim_{\b \to \ii} \oint_{\vp(0,\vec x) = \vp(\b,\vec x)} \cald \phi\,  e^{-S[\vp (t,\vec x)]+ \vJ \cd \vp(T)} \d\lp \vbp(0)-\v u\rp.
\ead 

The above expression has the interpretation as the probability of $\vbp(0)$ having the value $\v u$ in the presence of the source term $\vJ \cd \vbp(T)$. Under the assumption that there is still a unique vacuum in the presence of an infinitesimal source term (that is, no spontaneous symmetry breaking), the vacuum expectation value of $\d\lp \vbp(0)-\v u\rp$ will change infinitesimally for an infinitesimal source, and the system observed at $t=0$ will continue to have a finite correlation length. 
Moreover, as the source is chosen to be spatially constant, the system will remain uniform in space. In the present discussion, it is assumed that the expectation value of $\ev{\vbp(0,\vec x)}$ vanishes. In the presence of an infinitesimal source term at time $T$, indicated by a subscript $J,T$, the expectation value may shift infinitesimally as $\ev{\vbp(0,\vec x)}_{J,T} =  W(T) \vec J$, where $W(T)$ is an undetermined $N \times N$ matrix. From $O(N)$ invariance\footnote{For $N \neq 2$, $SO(N)$ invariance is enough to arrive at this conclusion. For $N = 2$, invariance under reflection needs to be assumed as otherwise $W_{ab}(T) \propto \e_{ab}$ is also a valid possibility where $\e_{ab}$ is the $2D$ Levi-Civita symbol.} it follows that $W(T) = \r(T) I$ where $I$ is the identity matrix and $\r(T)$ is a real function. It follows that the results of Sec. \ref{sec:LargeTime} can be used for the shifted fields $\vec{ {\phi '}}(0,\vec x) = \vec{\phi}(0,\vec x)-W(T) \vJ$. Therefore, from Eq. \eqref{eq:Pu} it follows that:

\bad 
P_{\vbp(0)}(\v u;\v J,T) = \ff{1}{\ss{(2\pi)^N \det \S(\v J,T)}}e^{-\ff 1 2 \lp \v u - W(T) \v J \rp^T \inv{\S(\v J,T)} \lp \v u - W(T) \v J \rp},
\label{eq:prob_J_u}
\ead
where $P_{\vbp(0)}(\v u;\v J,T)$ is the probability that $\vbp(0)$ is equal to $\v u$ in the presence of source $\v J$ which is inserted at time $T$.

The matrix $\S(\vec{J},T)$  transforms in the adjoint representation of $O(N)$: 
\bad 
\S(R\v J,T) = R \S(\v J,T) R^T,
\label{eq:SigmaR}
\ead
and consequently, the linear term in the expansion of $\S(\vec J,T)$ in $\vec J$ vanishes\ (the only invariant tensor of the $O(N)$ group with two indices is $\d_{ab}$). It follows that, to first order in $\v J$, one can replace $\S(\v J ,T)$ by $\S = \s^2 I$ where $\s$ is defined by Eq. \eqref{eq:sigma_O(N)_def}. Therefore, $P_{\vbp(0)}(\v u;\v J,T)$ can be expanded in $\v J$ as
\bad 
P_{\vbp(0)}(\v u;\v J,T) &= P^{(0)}_{\vbp(0)}(\v u) - \rho(T)\lp \v J \cd \nn_{\v u} \rp P^{(0)}_{\vbp(0)}(\v u)  + \co \lp J^2 \rp \\
&= P^{(0)}_{\vbp(0)}(\v u) \lp 1 + \rho(T)\lp \v J \cd \inv \S \v u \rp \rp  + \co \lp J^2 \rp \\ 
&= P^{(0)}_{\vbp(0)}(\v u) \lp 1 + \ff{\rho(T)}{\s^2}\vec J \cd \vec u \rp   + \co \lp J^2 \rp.
\ead  
Defining $\rho = \lim_{T \to -\ii} e^{-mT} \rho(T)$ (which is finite since $\rho(T)\propto e^{mT}$ for $T\to-\infty$), from Eqs. \eqref{eq:m_to_phi} and \eqref{eq:phi_to_J} one obtains:
\bad 
\mel{\O}{\d \lp \vbp(0)-\v u\rp}{m,a} = \ff{\r}{\s^2} P^{(0)}_{\vbp (0)}(\v u) u_a\,.
\ead 
It follows that the first order correction  to the joint distribution function (arising from the lowest energy excited multiplet of states) is:
\bad
P^{(1)}_{\vbp(t),\vbp(0)}(\v u,\v v) = e^{-mt}\D \v u \cd \v v P^{(0)}_{\vbp(t),\vbp(0)}(\v u, \v v)\,,
\label{eq:P1uv}
\ead 
where $\D = \ff{\r^2}{\s^4}>0$. 
Therefore, including the first order correction, the full joint distribution is given by:

\bad 
P_{\vbp (t),\vbp (0)}(\v u , \v v) &= \ff{1}{\lp 2\pi \rp^N \s^{2N}}e^{-\ff{1}{2\s^2}\lp \v u \cd \v u + \v v \cd \v v \rp} {\lp 1 + e^{-mt}\D \v u \cd \v v \rp} + \co \lp e^{-m't} \rp
\\
&= \ff{1}{\lp 2\pi \rp^N \s^{2N}}e^{-\ff{1}{2\s^2}\lp \v u \cd \v u + \v v \cd \v v \rp + e^{-mt}\D \v u \cd \v v } + {\co \lp e^{-m't} \rp},
\label{eq:Pjoint_corr}
\ead
where $2m \geq m' > m$, and in the second line the correction term has been exponentiated.
It follows that for $N=2$ the results of the Sec. \ref{sec:Free} are valid up to the first order in $K_f(t)$ with the identification $K_f(t) \equiv e^{-mt}\D\s^4$. 

In a similar fashion, one obtains the first order correction to $P_{C_{\op}}(t)$ in Eq.~\eqref{eq:PCdot} as
\bad 
P^{(1)}_{C_{\op}(t)}(x) &= \ff{e^{-mt}\D }{2\pi }\int d\o e^{i\o x} \int d\v u d\v v e^{-i \o \v u \cd \v v}\v u \cd \v v P^{(0)}_{\vbp(t),\vbp(0)}(\v u, \v v) \\ 
&=\ff{e^{-mt}\D }{2\pi}\int d\o e^{i\o x} \lp i \ff{\pp}{\pp \o}\rp \int d\v u d\v v e^{-i \o \v u \cd \v v} P^{(0)}_{\vbp(t),\vbp(0)}(\v u, \v v) \\ 
&= \ff{1}{2\pi}\int d\o e^{i \o x}\lp i e^{-mt}\D \ff{\pp}{\pp \o}\rp \Phi^0_{C_{\op}(T)}(\o).
\ead 
\ewt
From the above relation, up to the first order correction it is seen that $\Phi_{C_{\op}(t)}(\o)$ is given as
\bad 
\Phi_{C_{\op}(t)}(\o) &= \Phi^{(0)}_{C_{\op}(t)}(\o) + i e^{-mt}\D \ff{\pp}{\pp \o}\Phi^{(0)}_{C_{\op}(t)}(\o) 
+ \co\lp e^{-m't}\rp \\ 
&= \Phi^{(0)}_{C_{\op}(t)}(\o+ie^{-mt}\D) + \co \lp e^{-m't}\rp.
\ead 
From this expression, it is clear that including the first order correction to the characteristic function $\Phi_{C_{\op}(t)}(\o)$ is equivalent to a shift in the argument in the imaginary direction that decreases as $t\to\infty$.

\subsection{All Times}
It has been shown in Eqs.~\eqref{eq:PSigmauv} and \eqref{eq:Pjoint_corr} that both in the large time limit and even including its first correction, the joint probability density $P_{\vbp(t),\vbp(0)}(\v u,\v v)$ is given by a coupled Gaussian probability density, in the infinite volume limit. In this subsection, it will be shown that this fact is valid including all corrections in the infintie volume limit.

Consider a division of the spatial volume into boxes of sizes $l>\xi$, where $\xi$ is the correlation length in lattice units. Boxes are enumerated by $I=1,\cds,K$ with each box having $E = \ff{L^d}{K}$ sites. For each box, box-averaged fields are defined as  

\bad 
\bb \phi_{a,I}(t) = \ff{1}{\ss{E}}\sum_{\vec x \sim I}\phi_a(t,\vec x),
\ead 
where the summation over $\vec x \sim I$ is over all  sites in the $I$th box. One notes immediately that
\bad 
\bb \phi_{a}(t) = \ff{1}{\ss K}\sum_{I}\bb \phi_{a,I}(t).
\ead 
For $I \neq J$, $\bb \phi_{a,I}(t)$ can be considered independent of $\bb \phi_{b,J}(t')$ up to the corrections proportional to $e^{-{\ss{l_{IJ}^2 + (t-t')^2}}/{\xi}}$ where $l_{IJ}$ is the distance between centers of the boxes $I$ and $J$ in lattice units. 

For each box $I$, a $2N$-tuple of fields
\be
\vec Q_I(t) = \bpm \bp_{1,I}(t) \\ \cds \\ \bp_{N,I}(t) \\ \rule{2cm}{0.4pt} \\ \bp_{1,I}(0) \\ \cds \\ \bp_{N,I}(0) \epm.
\ee
will be considered.
The joint distribution of $\vbp(t)$ and $\vbp(0)$ is then equivalent to the distribution of $\vec Q(t)$ where $\vec Q(t)$ is defined by
\bad 
\vec Q(t) = \ff{1}{\ss K}\sum_{I=1}^K \vec Q_I(t).
\ead 

The goal here is to show that this vector follows a Gaussian probability distribution.
The main assumption needed to show this is the independence of $\vec Q_I(t)$ and $\vec Q_J(t)$ for $I \neq J$. The errors due to this approximation can be made arbitrarily small in the infinite volume limit by taking $l, K \to \ii$. Therefore, in the infinite volume limit, one may write
\bad 
\lim_{V \to \ii}P_{\vec Q(t)}(\vec z) &= \lim_{K \to \ii}\int \prod_{I=1}^K d\vec z_I \d \lp \vec z - \ff{1}{\ss K}\sum_{J=1} \vec z_J \rp  P_{\vec Q_I(t)}(\vec z_I) \\ 
&= \ff{1}{(2\pi)^{2N}} \lim_{K \to \ii} \ff{1}{\lp 2\pi \rp^{2NK}}\int d\vec w e^{i \lp \vec z - \ff{1}{\ss K}\sum_{J=1} \vec z_J \rp \cd \vec w}\prod_{I=1}^K d\vec z_I d \vec w_I \Phi_{\vec Q_1(t)}(\vec w_I)e^{i \vec w_I \cd \vec z_I}\\ 
&= \ff{1}{(2\pi)^{2N}}  \int d\vec w e^{i \vec w \cd \vec z} \lim_{K \to \ii}\prod_{I=1}^K \Phi_{\vec Q_1(t)} \lp \ff{\vec w}{\ss K} \rp \\ 
&= \ff{1}{(2\pi)^{2N}}  \int d\vec w e^{i \vec w \cd \vec z}\lim_{K \to \ii}\lp \Phi_{\vec Q_1(t)} \lp \ff{\vec w}{\ss K} \rp \rp^K \\ 
&=\ff{1}{(2\pi)^{2N}}  \int d\vec w e^{i \vec w \cd \vec z}\lim_{K \to \ii}  \exp{-\ff{1}{2K} \vec w^T \hat \Sigma(t) \vec w + \co \lp \ff{1}{K\ss K}\rp}^K \\ 
&=\ff{1}{(2\pi)^{2N}}\int d\vec w e^{i \vec w \cd \vec z}e^{-\ff 1 2 \vec w^T \hat \Sigma(t) \vec w} \\ 
&= \ff{1}{(2\pi)^N \ss{\det \hat \Sigma(t)}}e^{-\ff 1 2 \vec z^T \inv{\hat \Sigma(t)}\vec z},
\label{eq:PQ}
\ead
where translational invariance has been used to set $\Phi_{\vec Q_I(t)} = \Phi_{\vec Q_1(t)}$ for all $I$. Note also that $\Phi_{\vec Q_1(t)}(0) = 1$ and $\Phi'_{\vec Q_1(t)}(0) = 0$ which permits one to write $\Phi_{\vec Q_1(t)}(\vec w) = \exp{-\ff 1 2 \lp \vec w \rp^T \hat \Sigma(t) \vec w + \co(w^3)}$. Here, 
\bad 
\hat \Sigma(t) = 
\begin{pmatrix}
  \Sigma
 & \ti \Sigma(t) \\
  \ti \Sigma(t) & 
  \Sigma
\end{pmatrix},
\ead 
where $\Sigma_{ab} = \s^2 \d_{ab}$ was introduced in Sec.~\ref{sec:LargeTime} and  $\ti \Sigma_{ab}(t) = \ev{\bp_{a,1}(t) \bp_{b,1}(0)}= \ev{\bp_a(t) \bp_b(0)}$.  From $O(N)$ invariance it follows that $\ti \S_{ab}(t) = K(t)\d_{ab}$ for some $K(t)$. Therefore, $\hat \Sigma(t)$ can be written in the form $\hat \Sigma_{(\tt a)(\tt' b)}(t) = \d_{ab}\bar \Sigma_{\tt \tt'}$ where $\tt \in \{1,2\}$ with different $\tt$ corresponding to different times ($\tau=1,2$ correspond to $0$ and $t$, respectively) and $\bar \Sigma_{\tt \tt'}(t)$ is given by
\bad 
\bar \Sigma(t) = \bpm \s^2 & K(t) \\ K(t) & \s^2 \epm.
\ead 
It follows that $\inv{\hat \S_{(\tt a )(\tt' b)}}(t) = \d_{ab} \inv{\bar \Sigma_{\tt \tt'}}$ where $\inv{\bar \Sigma_{\tt \tt'}}$ is given as
\bad 
    \inv{\bar \S}(t) = \ff{1}{\s^4- K(t)^2}\bpm \s^2 & -K(t) \\ -K(t) & \s^2 \epm.
\ead 
Therefore, the joint probability distribution of $\vbp(t)$ and $\vbp(0)$ in the limit $L\to\infty$ is given from Eq.~\eqref{eq:PQ} as
\bad 
P_{\vbp(t),\vbp(0)}(\v u,\v v) &= \ff{1}{(2\pi)^N \lp \s^4 - K(t)^2 \rp^\ff{N}{2}}  \exp{-\ff{\s^2\lp  |\vec u|^2 + |\vec v|^2 \rp}{2\lp \s^4 - K(t)^2 \rp} + \ff{K(t)\lp \v u \cd \v v \rp}{\s^4-K(t)^2}},
\label{eq:jointN}
\ead 
Note that this expression has been derived without taking $t\to\infty$ and reduces to Eq.~\eqref{eq:free_joint_pdf} for $N=2$.

\subsection{Distribution of $C_{\op}(t)$ }
\label{sec:Cdotdist}
For arbitrary $t$ and $\b$, it follows from Eq. \eqref{eq:jointN} that $P_{C_\op}(x)$ is given by
\bad 
P_{C_\op{(t)}}(x;\omega_+,\o_-,N) &= 
\int du dv \delta(x-\vec{u}\cdot\vec{v}) P_{\vbp(t),\vbp(0)}(\v u,\v v)
\\
&=\ff{1}{2\pi}\int_{-\ii}^{\ii}d\o e^{i\o x}\ff{\lp \o_+ \o_-\rp^{\ff N 2}}{\lp \o - i\o_+\rp^{\ff N 2}\lp \o + i\o_- \rp^{\ff N 2}},
\label{eq:Ixww_def}
\ead 
where the poles $\o_\pm$ are given by
\bad 
\o_{\pm} = \ff{1}{\s^2 \pm K(t)},
\ead 
and the dependence on $N$ is made explicit.
Note that Eq.~\eqref{eq:Ixww_def} defines 
\be
\Phi_{C_\op(t)}(\o)=\ff{\lp \o_+ \o_-\rp^{\ff N 2}}{\lp \o - i\o_+\rp^{\ff N 2}\lp \o + i\o_- \rp^{\ff N 2}}.
\ee

The integral in Eq.~\eqref{eq:Ixww_def} will be calculated explicitly for $x>0$; the result for  $x<0$ can be obtained  using the relation $P_{C_\op{(t)}}(x;\o_+,\o_-,N) = P_{C_\op{(t)}}(-x;\o_-,\o_+,N)$. 
The calculation begins with the observation that $P_{C_\op{(t)}}(x;\o_+,\o_-,N)$ is an analytic function of $N$ for $Re(N)>0$. This can be seen by using the relation $e^{i\o x} = -\ff{i}{x}\pp_\o e^{i\o x}$ to find another expression for $P_{C_\op{(t)}}(x;\o_+,\o_-,N)$ through integration by parts:
\bwt 
\bad 
P_{C_\op{(t)}}(x;\o_+,\o_-,N) &= \ff{-i\lp \o_+ \o_-\rp^{\ff N 2}}{2\pi x} \lk \ff{e^{i\o x}}{\lp \o - i\o_+\rp^{\ff N 2}\lp \o + i\o_- \rp^{\ff N 2}} \rk^{\o=\ii}_{\o=-\ii} \\ 
&\qquad + \ff{iN\lp \o_+ \o_-\rp^{\ff N 2}}{4\pi x}\int_{-\ii}^{\ii}d\o \ff{\lp 2\o - i\lp \o_+ - \o_-  \rp \rp e^{i\o x}}{\lp \o - i\o_+\rp^{1+\ff N 2}\lp \o + i\o_- \rp ^{1+\ff N 2}}.
\ead

\begin{figure}[!t]
\centering
\begin{tikzpicture}[thick]
  \node[scale=1.2] at (2.7,3.0) 
    {$\o$-plane};
  % Axes:
  \draw [->] (0,0) -- (-1.05*\xr,0) ;
  \draw [->] (0,0) -- (1.05*\xr,0) node [right] {Re};
  \draw [->,decorate,decoration={zigzag,segment length=4,amplitude=1,pre=lineto,pre length=15,post=lineto,post length=3}] (0,0) -- (0,-1.05*\yr) ;
  \draw [->,decorate,decoration={zigzag,segment length=4,amplitude=1,pre=lineto,pre length=15,post=lineto,post length=3}] (0,0) -- (0,1.05*\yr) node [above] {Im};
  \draw[yshift=9.8,blue!60!black,decoration={markings,mark=between positions 0.125 and 0.875 step 0.25 with \arrow{>}},postaction={decorate}] (-\xr/20,9*\yr/10) -- (-\xr/20,\yr/20) arc (-180:0:\xr/20) (\xr/20,\xr/20) -- (\xr/20,9*\yr/10) node[below right] {$C$};
  \filldraw [black] (0,0.45) circle (1.3pt)  node [right] {$\ \omega_+$};
  \filldraw [black] (0,-0.45) circle (1.3pt)  node [right] {$\ \omega_-$};
\end{tikzpicture}
\caption{The integration contour used to evaluate $P_{C_\op}(x;\o_+,\o_-,N)$ for $x>0$.}
\label{fig:contour}
\end{figure}
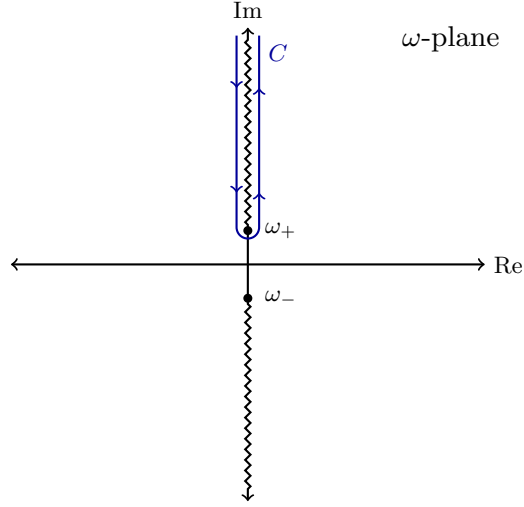
The analyticity of $P_{C_\op{(t)}}(x;\o_+,\o_-,N)$ for $Re(N)>0$ then follows because the first term above vanishes for $Re(N>0)$ and the integral in the second term is absolutely convergent for $Re(N)>0$.

Assuming $x>0$, one can deform the integration contour of the integral given in Eq. \eqref{eq:Ixww_def} to the upper complex plane, see Fig. \ref{fig:contour}. For $0<Re(N)<2$, the contribution of the integral over the semi-circle below $\omega_+$ vanishes as the radius of the semi-circle goes to $0$. Therefore, for $0 < Re(N) < 2$, 
%\bwt
\bad 
P_{C_\op{(t)}}(x>0;\omega_+,\o_-,N) &= \ff{\lp \o_+ \o_-\rp^{\ff N 2}}{2\pi}2 \sin{\ff{\pi N}{2}} \int_{\o_+}^\ii dy \ff{\exp{-y x}}{\lp y-\o_+ \rp^{\ff N 2} \lp y+\o_- \rp^{\ff N 2}} \\ 
&= e^{\ff 1 2 (\o_--\o_+) x }\lp \ff{\o_+ \o_-}{\o_+ + \o_-}\rp^{\ff N 2}\ff{\ss{\o_+ + \o_-}}{\ss{\pi}\Gamma(\ff N 2)} x^{\ff{N-1}{2}} K_{\ff{N-1}{2}} \lp \ff{1}{2}(\o_++\o_-)x\rp ,
\ead 
{where as before $K_n(x)$ is a modified Bessel function. Since this expression is analytic for $\Re(N)>0$, it can be continued to all $\Re(N)>0$.}

Imposing the above relation between $x>0$ and $x<0$, the full distribution can be expressed compactly as: 
\bad 
P_{C_\op(t)}(x;\omega_+,\o_-,N) 
&= e^{\ff 1 2 (\o_--\o_+) x }\lp \ff{\o_+ \o_-}{\o_+ + \o_-}\rp^{\ff N 2}\ff{\ss{\o_+ + \o_-}}{\ss{\pi}\Gamma(\ff N 2)} \abs{x}^{\ff{N-1}{2}}  K_{\ff{N-1}{2}} \lp \ff{1}{2}(\o_++\o_-)\abs{x}\rp. 
\label{eq:shifted_func}
\ead
\ewt
For large times $K(t) \to 0$ and  $\o_\pm \to \ff{1}{\s^2}$  in Eq.~\eqref{eq:shifted_func}, so it follows that
\bad 
P^{(0)}_{C_{\op}(t)}(x;\s,N) &= \lim_{t\to \ii, \b \to \ii} P_{C_{\op}(t)}(x;\s^2) \\
&= \ff{2^{\ff{1-N}{2}}\s^{-N-1}}{\Gamma \lp \ff N 2 \rp \ss \pi}\abs{x}^{\ff{N-1}{2}}K_{\ff{N-1}{2}}\lp \ff{\abs{x}}{\s^2} \rp.
\label{eq:asym_func}
\ead
For $N=1$, Eq.~\eqref{eq:asym_func} reduces to the result found previously in Ref.~\cite{Yunus:2022pto}.

\section{Numerical Analysis for the $O(2)$ Model}

\label{sec:numerical}
In this section, the relations derived in Sec. \ref{Sec:ON} will be tested numerically for the case $N=2$ in two dimensions. The action of the theory is given\footnote{For these calculations, an adaptation of the publicly available code at \cite{agimenezromero} is used, where periodic boundary conditions are assumed and heatbath algorithm is used to generate configurations.} by
\bad 
S = \sum_i \lk \lp 2-\ff \theta 2\rp \abs{\psi_i}^2 + \ff \chi 4 \abs{\psi_i}^4 \rk - \ff 1 2 \sum_{\lc i j \rc} \psi_i^\star \psi_j + \psi_j^\star \psi_i,
\ead 
where $i$ labels the sites, $\sum_{\lc i j \rc}$ indicates summation over all pairs of neighboring points, $\theta$ and $ \chi$ are the couplings and $\psi_i$ is a complex field (equivalent to the $O(2)$ model). The simulations are performed for various geometries $L\times \beta$ and for various sets of couplings that are parameterized by a single parameter $s$ through the relations
\bad
\chi &= - \ln{s}, \\  
\theta &= - \ln \lp 1-s \rp.
\label{Eq:couplings}
\ead 

To demonstrate the validity of the various results introduced in Sections \ref{sec:Free} and \ref{Sec:ON}, and in particular to test the assumptions made in deriving them, it is useful to consider the total variation $\calt$ between the empirical distribution $E(q)$ determined from the numerical calculations and any proposed probability distribution $P(q)$:
\bad 
\calt = \ff 1 2 \int dq\, \abs{E(q)-P(q)}.
\ead 
This quantity is unity if the distributions have support on disjoint domains and vanishes when the distributions are identical. Results will be presented for $C_{\op}(t)$, so comparisons are made to the distributions $P^{(0)}_{C_{\op}}(t)$ valid at asymptotically large times (Eq. \eqref{eq:asym_func}) and to the improved distribution $P_{C_{\op}}(t)$  (Eq. \eqref{eq:shifted_func}) that incorporates  corrections to the asymptotic case. The asymptotic distribution depends on a single parameter ($\sigma$) and the improved distribution depends on two parameters ($\omega_\pm$). In comparing the empirical distributions to these analytic forms, the following estimators for $\o_\pm$ and $\s$ are used. Assume that one has $\cal N$ samples $x_i$ of $C_{\op}(t)$, where $i \in \{1,\cds,{\cal N}\}$. Then, estimators for $\o_\pm$ and $\s$ are given\footnote{The validity of these estimators, in the infinite sample size limit, follows from Eq. \eqref{eq:C_2d_dists}, noting that $P^{(0)}_{C_{\op}(t)}(x)$ is equivalent to $P_{C_{Im}(t)}(x)$ with the identification $\o_I(t) = \ff{1}{\s^2(t)}.$} as:
\bad
\hat \o_+ &= \frac{1-\ss{\frac{\hat{\abs{x}}-\hat{x}}{\hat{\abs{x}}+\hat{x}}}}{\hat x}, \\ 
\hat \o_- &= \frac{\ss{\frac{\hat{\abs{x}}+\hat{x}}{\hat{\abs{x}}-\hat{x}}}-1}{\hat x}, \\
\hat \s &= \sqrt{\hat{\abs{x}}},
\ead
where $\hat x$ and $\hat{\abs{x}}$ are defined by:
\bad 
\hat x &= \frac{1}{\cal N}\sum_i x_i,\\
\hat{\abs{x}} &= \frac{1}{\cal N}\sum_i \abs{x_i}.
\ead 

\begin{figure}[!t]
 	\begin{subfigure}{0.5\linewidth}
    	\includegraphics[width=\columnwidth]{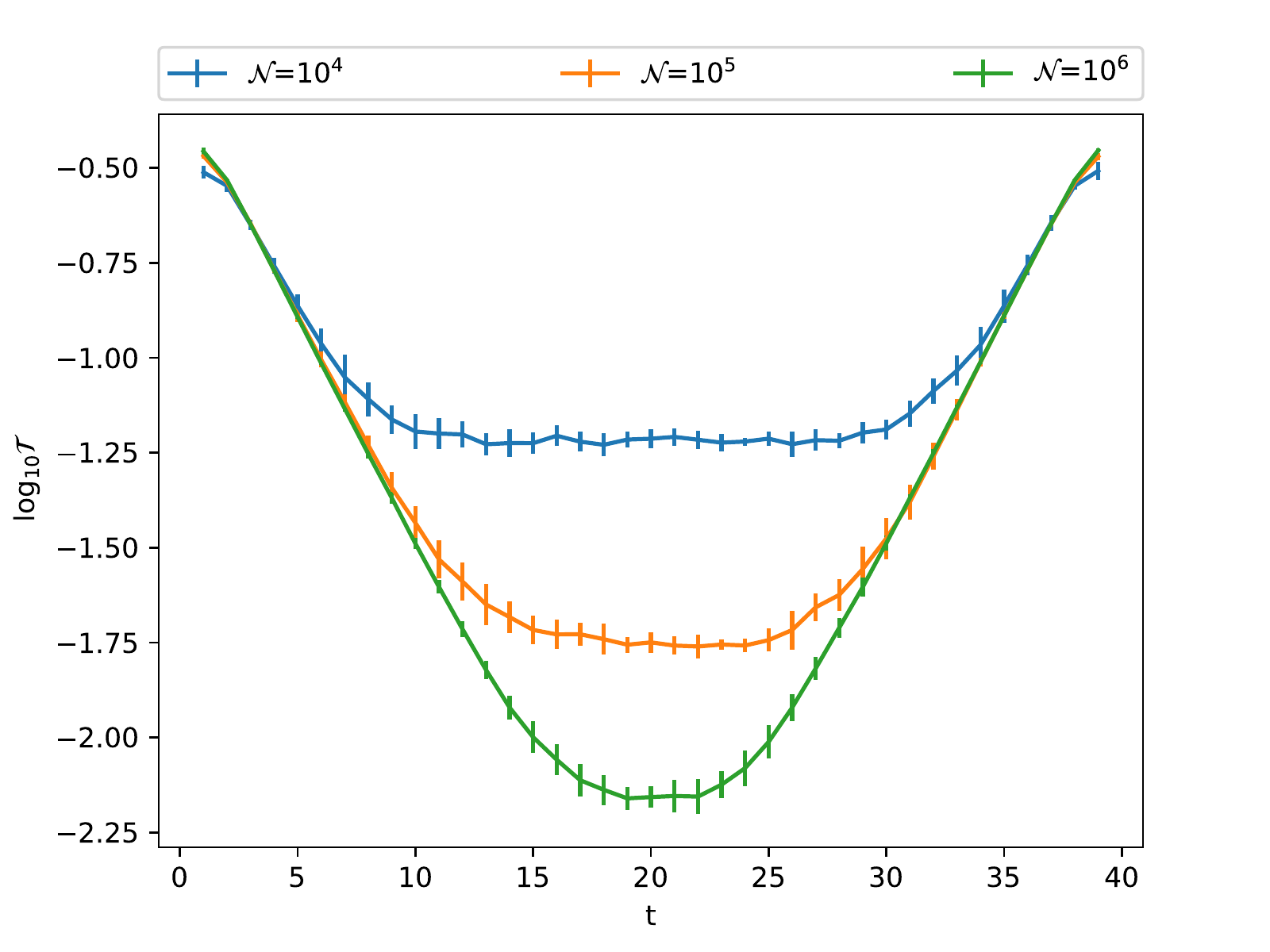}
    	\newsubcap{(a)}
	\end{subfigure}
	\begin{subfigure}{0.5\linewidth}
	\includegraphics[width=\columnwidth]{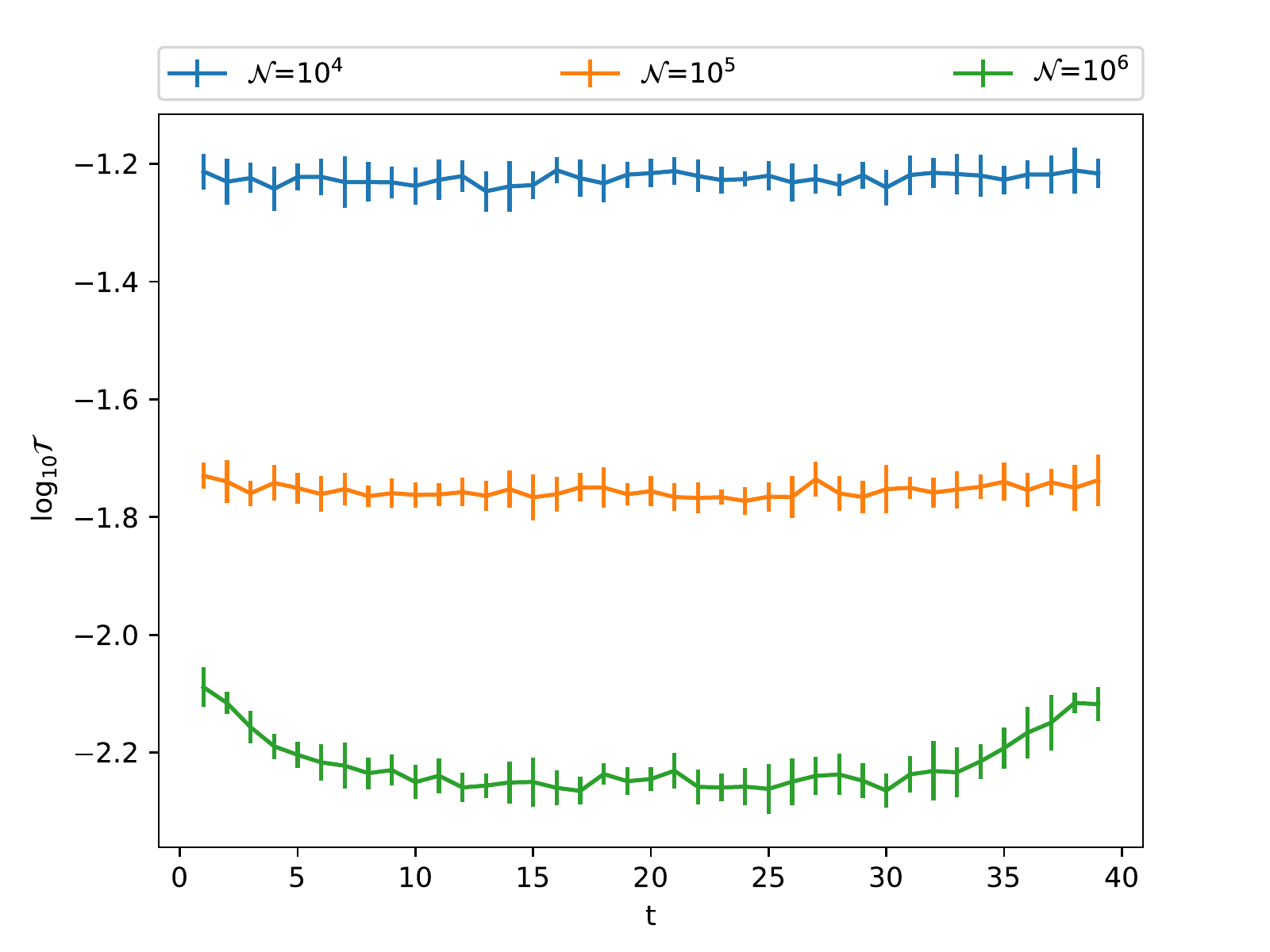}
	\newsubcap{(b)}
	\end{subfigure}
 \caption{The logarithm of the total variation as a function of temporal separation for three different sample sizes, $\caln$, for $s=0.56, L =100,\ \beta=40$ for the large time limit in \capa and with the corrections in \capb.}
 \label{Fig:SampleSize}
\end{figure}
In Fig. \ref{Fig:SampleSize}, the total variation is shown for a fixed geometry and  couplings as a function of the temporal separation of the operators in the correlation function for various values of the  number of samples used in the numerical calculations. Comparisons to both the asymptotic and improved distributions are shown. Since periodic temporal boundary conditions are used, the results are approximately symmetric around the midpoint of the temporal extent. As can be seen, the sample size sets a lower floor on the total variation in both cases but results for ${\cal N}=10^6$ samples are sufficient to cleanly resolve deviations from the asymptotic distribution, with that deviation achieving its minimum around the temporal midpoint. For the improved distribution, the total variation has not saturated even at ${\cal N}=10^6$ and should be viewed as an upper bound on the true total variation.  

\begin{figure}[!t]
 	\begin{subfigure}{0.5\linewidth}
    	\includegraphics[width=\columnwidth]{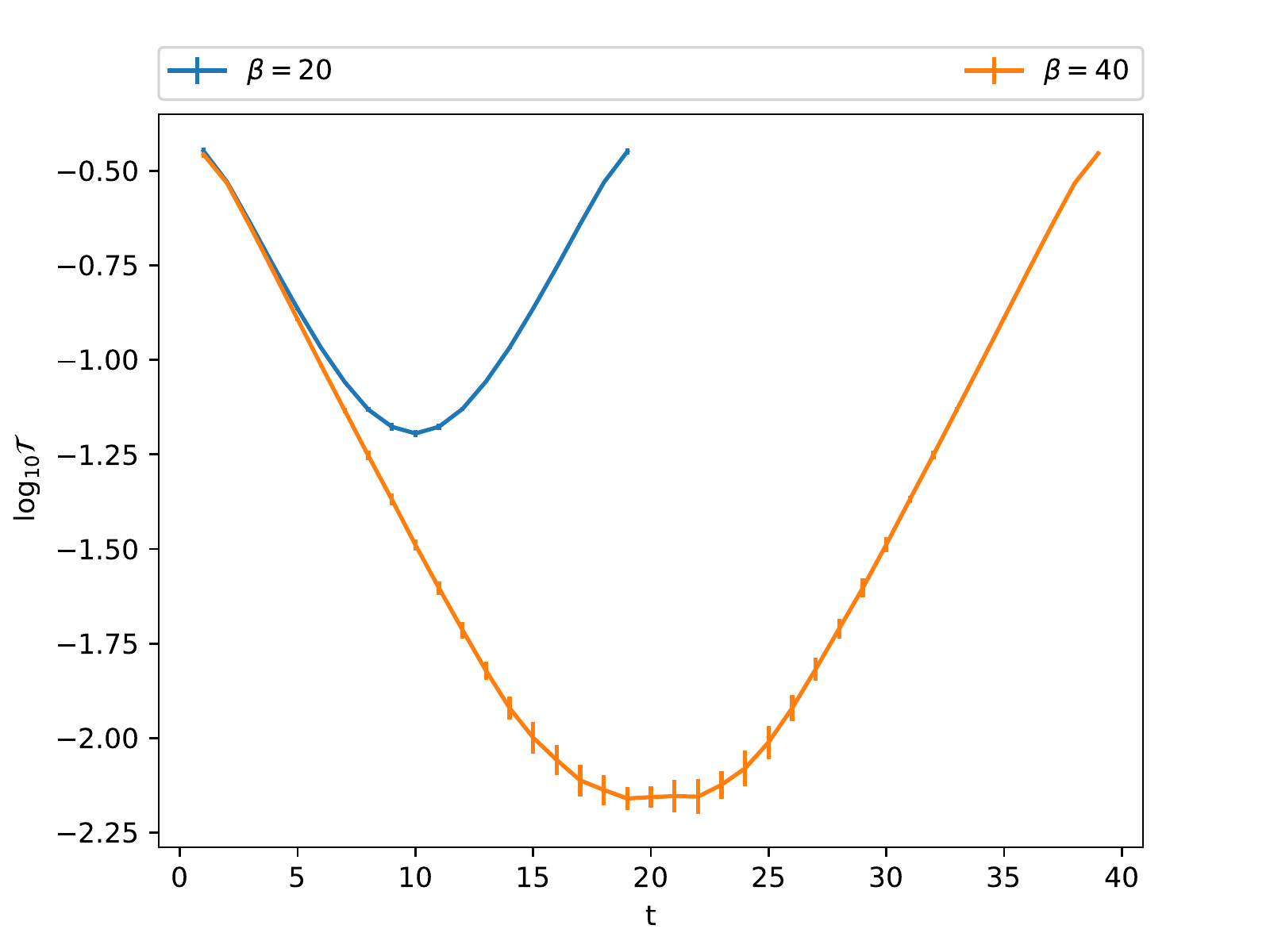}
    	\newsubcap{(a)}
	\end{subfigure}
	\begin{subfigure}{0.5\linewidth}
	\includegraphics[width=\columnwidth]{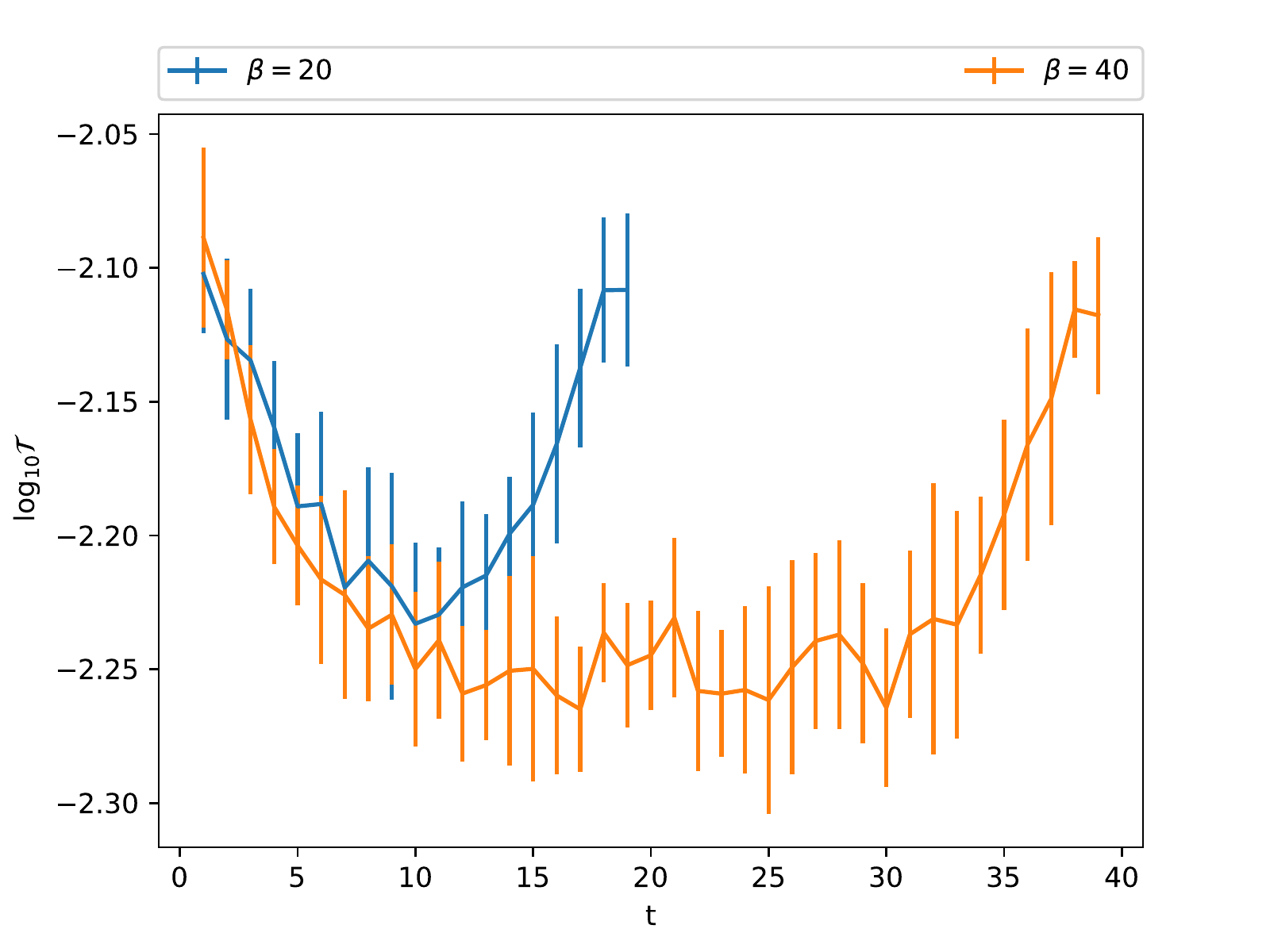}
	\newsubcap{(b)}
	\end{subfigure}
\caption{The logarithm of the total variation as a function of correlation function  time for two choices of the temporal extent, $\beta$,  with fixed  values of $s=0.56$, $L =100$, $\caln = 10^6$ for the large time limit in \capa and with the corrections in \capb.
}
\label{Fig:T}
\end{figure}

\begin{figure}[!t]
 	\begin{subfigure}{0.5\linewidth}
    	\includegraphics[width=\columnwidth]{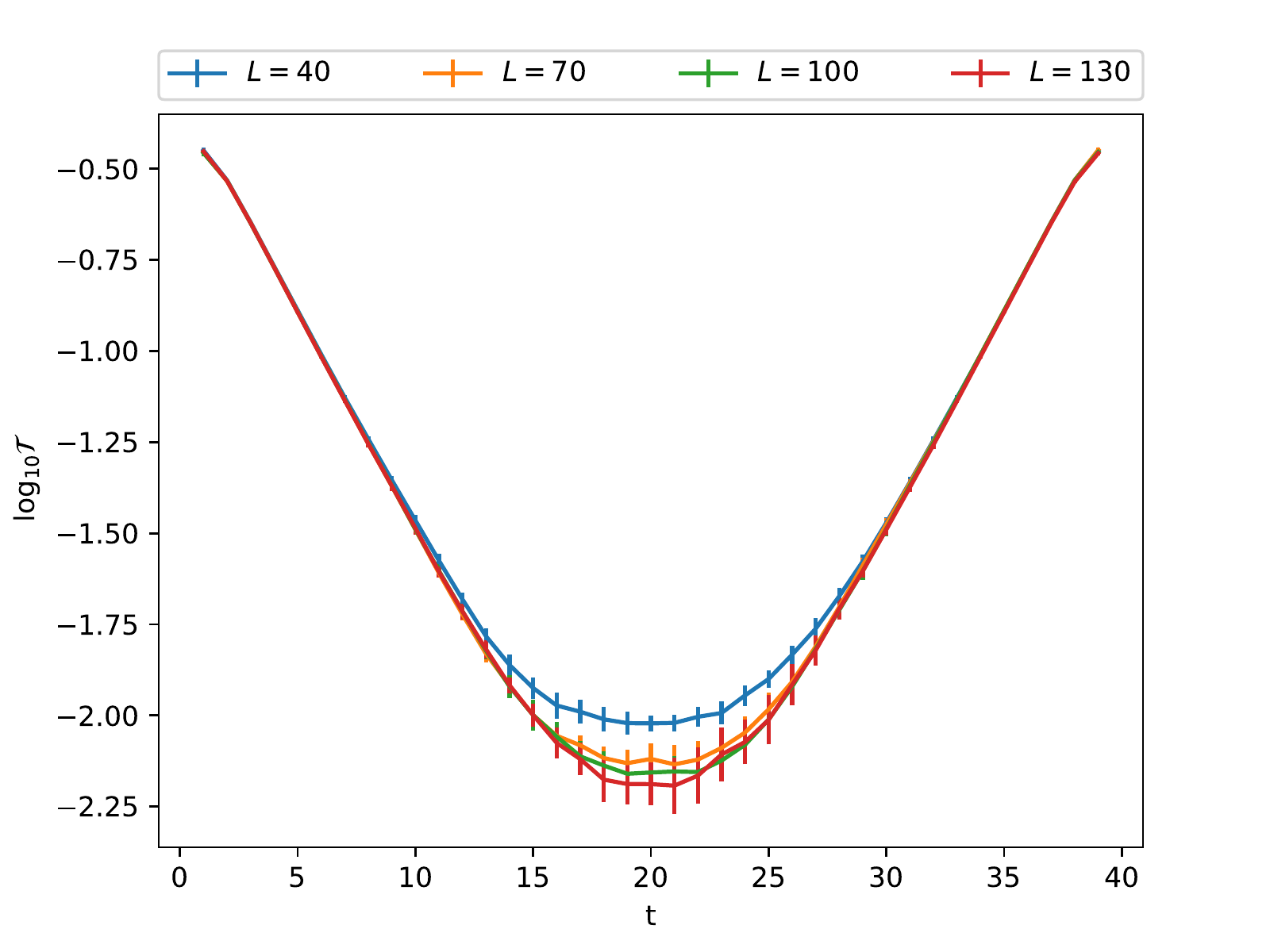}
    	\newsubcap{(a)}
	\end{subfigure}
	\begin{subfigure}{0.5\linewidth}
	\includegraphics[width=\columnwidth]{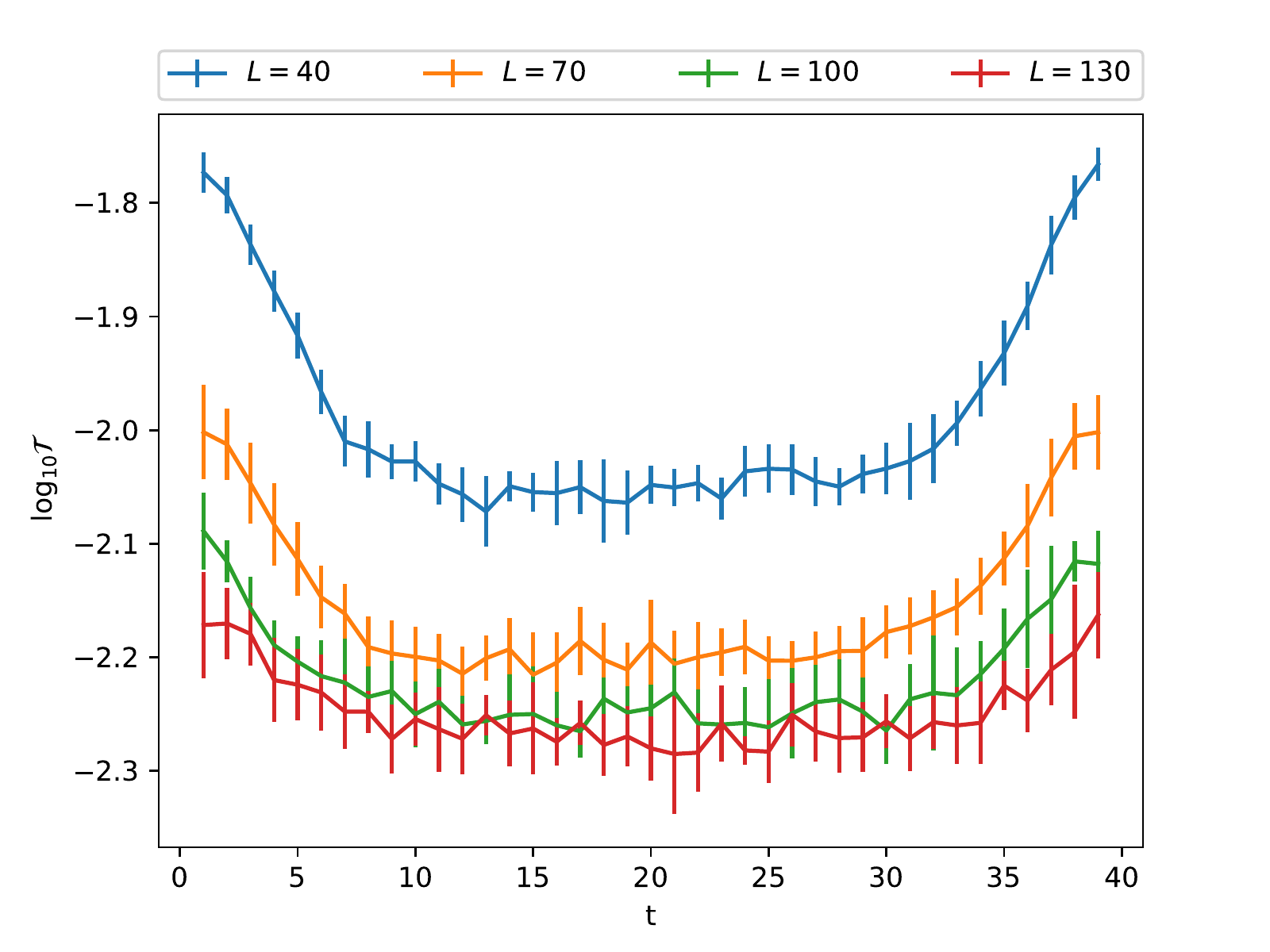}
	\newsubcap{(b)}
 \end{subfigure}
 \caption{The logarithm of the total variation as a function of correlation function  time for four values of the spatial extent, $L$,   with a fixed choice of  $s=0.56$,\ $\beta=40$,\ $\caln = 10^6$ for the large time limit in \capa and with the corrections in \capb.}
\label{Fig:L}
\end{figure}
Figure \ref{Fig:T} shows the dependence of the total variation on the temporal extent of the lattice geometry for fixed spatial extent and a single choice of the couplings. Similarly, Fig.~\ref{Fig:L} shows the dependence of the total variation on the spatial lattice extent for a fixed set of couplings and temporal extent. In both figures, the total variation is shown in comparison to the asymptotic and improved probability distributions. 
As can be seen in Fig.~\ref{Fig:T}, the temporal extent of the lattice geometry significantly affects the total variation, with the periodicity requirement competing against the approach of the correlation function to the asymptotic distribution. Even for  the improved distribution, deviations of the total variation are statistically resolved when the correlation function is measured for short time separations (including the effects of periodicity).
The behaviour of the total variation with respect to the spatial volume seen in Fig.~\ref{Fig:L} is in agreement with the 
discussions in the previous sections. As the spatial volume increases for a fixed $t$, the empirical distributions of the correlation function approach the analytic forms derived above assuming the infinite volume limit. For the couplings chosen in Fig.~\ref{Fig:L}, $L=100$ is sufficient to see agreement with the  asymptotic result, at least to the precision allowed by the finite sample size.

\begin{figure}[!t]
 	\begin{subfigure}{0.5\linewidth}
    	\includegraphics[width=\columnwidth]{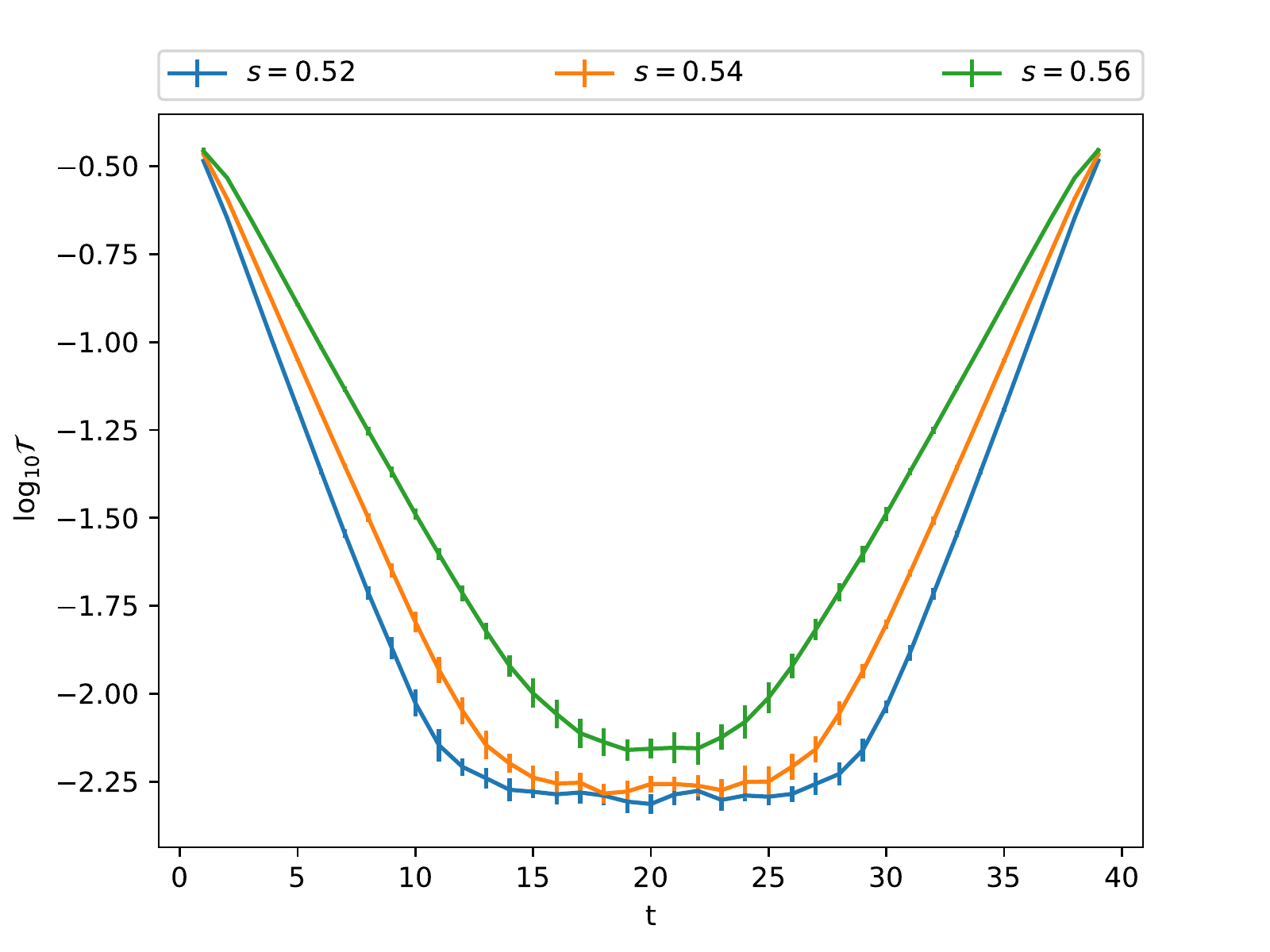}
    	\newsubcap{(a)}
	\end{subfigure}
	\begin{subfigure}{0.5\linewidth}
	\includegraphics[width=\columnwidth]{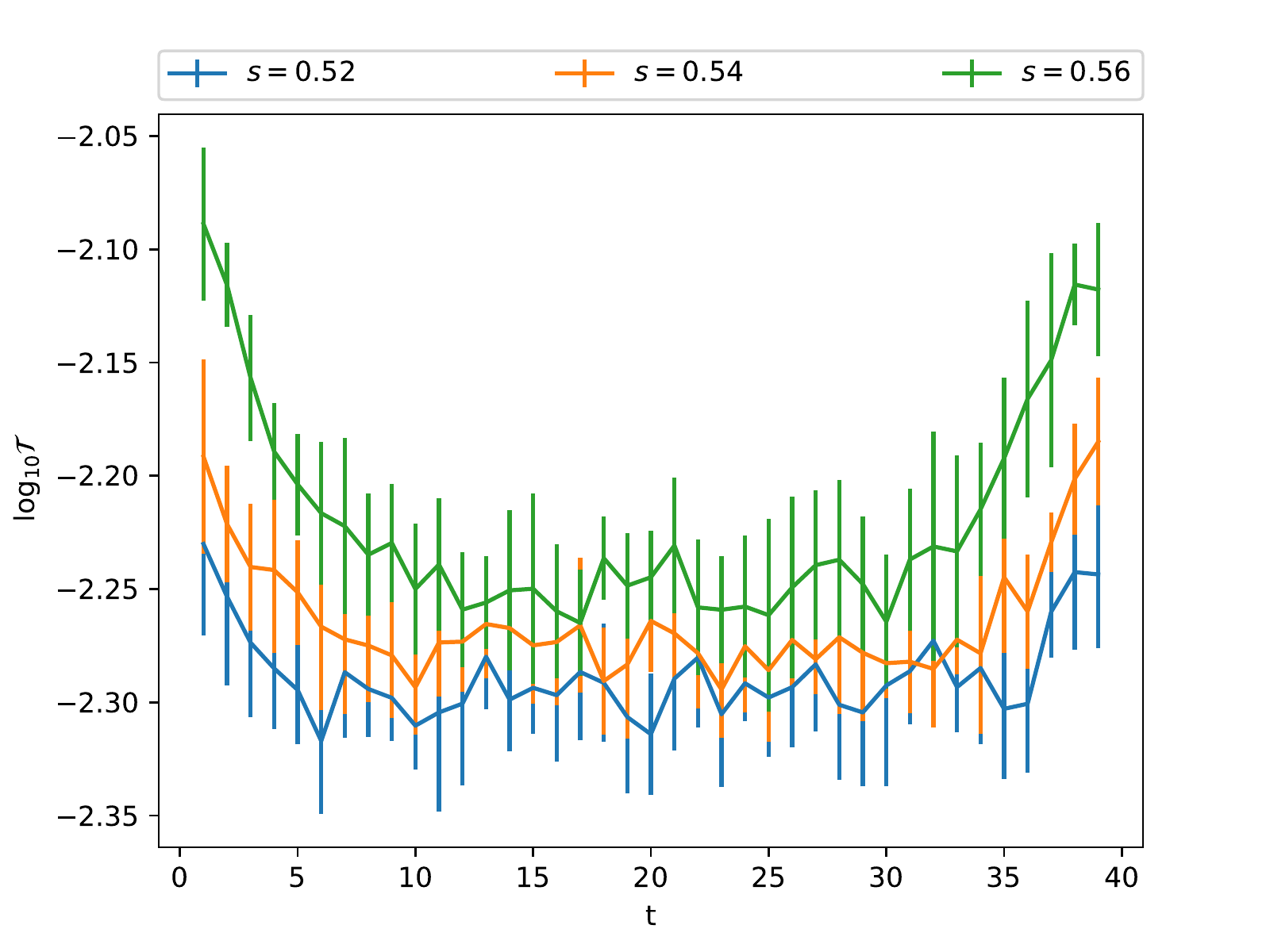}
	\newsubcap{(b)}
	\end{subfigure}
\caption{The logarithm of the total variation as a function of the temporal separation for various values of the parameter $s$ that determines the couplings. Fixed values of $L =100,\ \beta= 40$, $\caln = 10^6$ are used and results are shown for the large time limit in \capa and with the corrections in \capb.}
\label{Fig:s}
\end{figure}
\begin{figure}[!t]
 	\includegraphics[width=0.5\columnwidth]{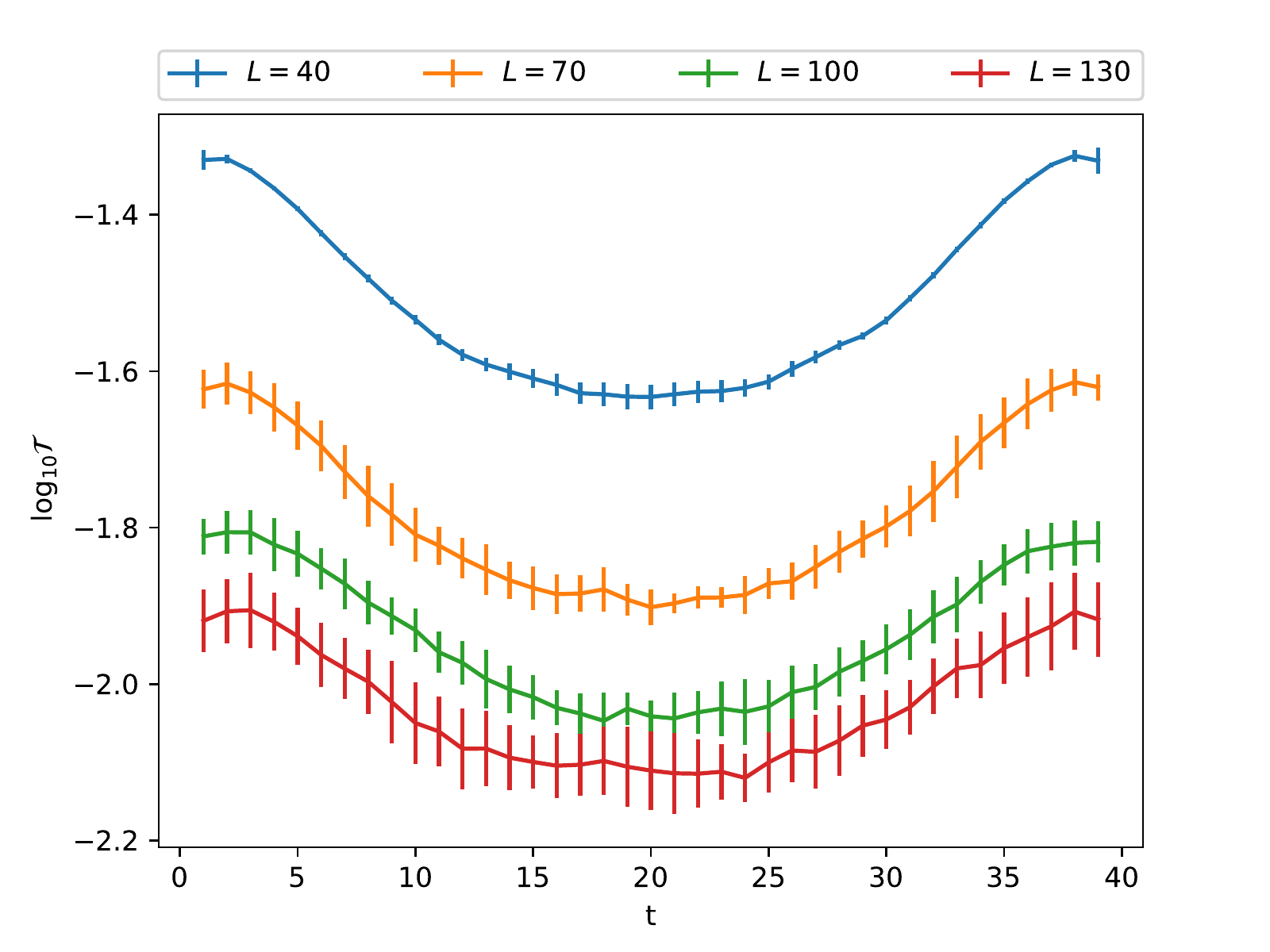}
\caption{The logarithm of the total variation with respect to the improved distribution as a function of the temporal separation in the correlation function for various different spatial extents $L$, for $\beta=40$,\ $\caln = 10^6$ and $s=0.59$.}
\label{Fig:notL}
\end{figure}
Figure \ref{Fig:s} show the variation of the total variation with respect to the parameter $s$ that determines the couplings through Eq.~\eqref{Eq:couplings}. As $s$ increases from $s=0$ towards $s\sim0.6$, the theory moves towards the  critical surface that separates the  quasi-ordered and disordered phases. As this surface is approached, the  correlation length diverges and as a consequence, the temporal separation of the correlation function and the spatial and temporal extents of the lattice geometry necessary to see agreement with the results in Sections \ref{sec:Free} and \ref{Sec:ON} (to a given accuracy) are correspondingly larger. As seen in  Fig.~\ref{Fig:s}, for larger values of $s$, corresponding to larger correlation lengths, the contamination from non-asymptotic contributions is more significant. This is particularly apparent for the total variation with respect to the asymptotic distribution, Eq.~\eqref{eq:asym_func}, but the trend is also seen in the deviation from the improved distribution, Eq.~\eqref{eq:shifted_func}.
A similar effect is seen comparing Fig.~\ref{Fig:notL} and Fig.~\ref{Fig:L}(b) which only differ in the value of $s$ that is used; clearly, for the case with the larger correlation length larger deviations from the improved distribution are seen.

The numerical results above summarised through calculations of the total variation  with respect to the asymptotic and improved distributions provide strong evidence for the  
validity of the assumptions that have been used to derive the results of the previous sections. 
To provide further support, Figs.~\ref{Fig:$s=0.52,L=40$}--\ref{Fig:$s=0.59, L=130$} show histograms of the empirical probability distributions for the real part of the correlation function  $P_{C_{\op}(t)}$ obtained for various couplings and geometries at representative temporal separations. As has been seen in studies of correlation function distributions in other contexts \cite{Beane:2009gs,DeGrand:2012ik,Wagman:2016bam,Wagman:2017gqi,Wagman:2017xfh,Detmold:2018eqd}, the distributions are strongly asymmetric under reflection about $x=0$ at small times but become increasingly symmetric as $t$ increases. Note that in all cases, the distributions have support for $x<0$ and so are not describable by log-normal distributions \cite{DeGrand:2012ik}. 

Figures~\ref{Fig:$s=0.52,L=40$}--\ref{Fig:$s=0.59, L=130$} also show the best fits for the asymptotic and improved analytic distributions in each case. As can be seen from the figures, for large enough spatial volume, the improved distribution accurately describes the histograms for all of the temporal separations that are presented. The asymptotic distribution (necessarily symmetric about $x=0$) provides a poor description at small times but the disagreement decreases as $t$ increases.
\begin{figure}
 	\begin{subfigure}{0.5\linewidth}
    	\includegraphics[width=\columnwidth]{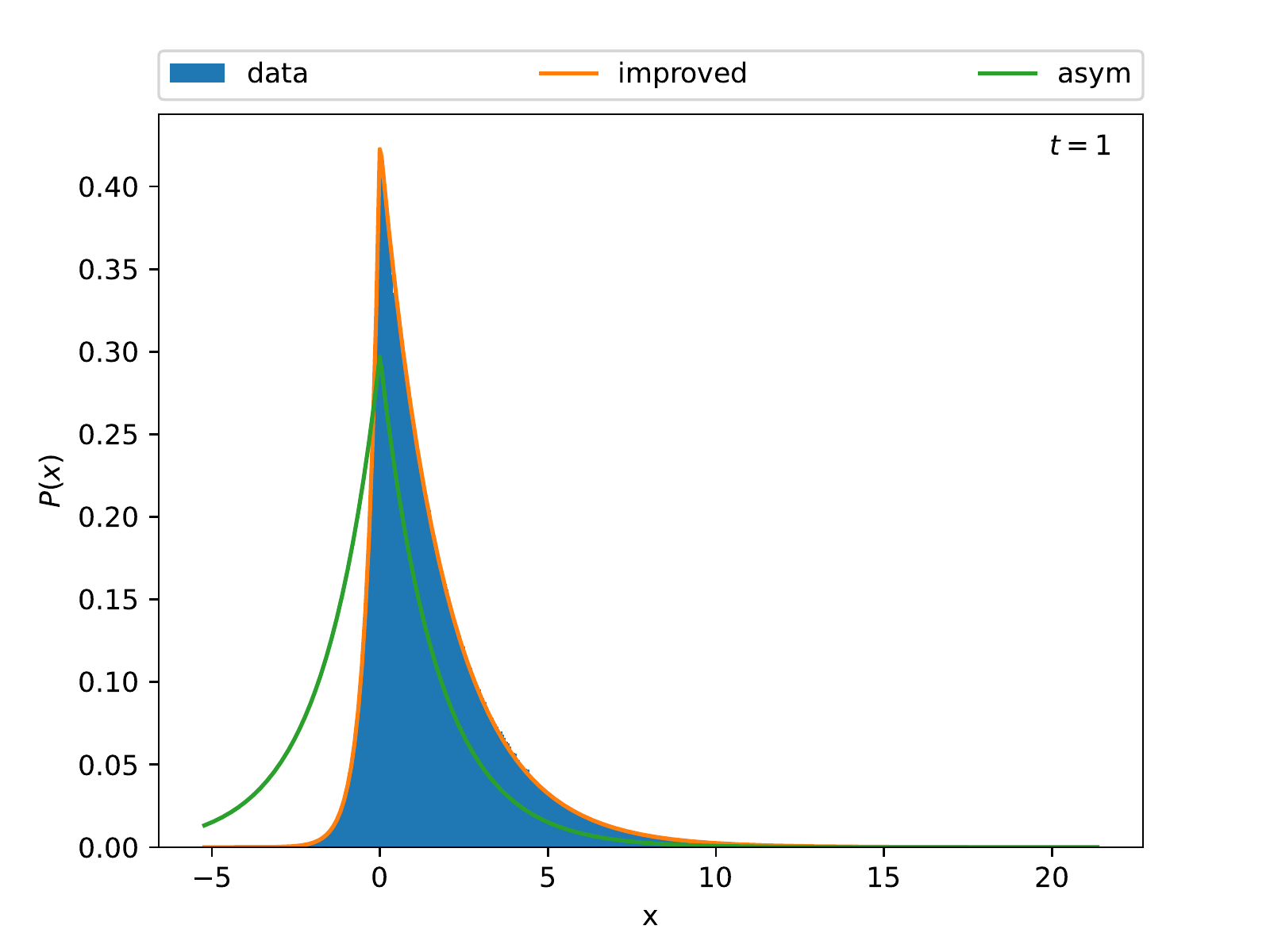}
    	\newsubcap{(a) }
	\end{subfigure}
	\begin{subfigure}{0.5\linewidth}
	\includegraphics[width=\columnwidth]{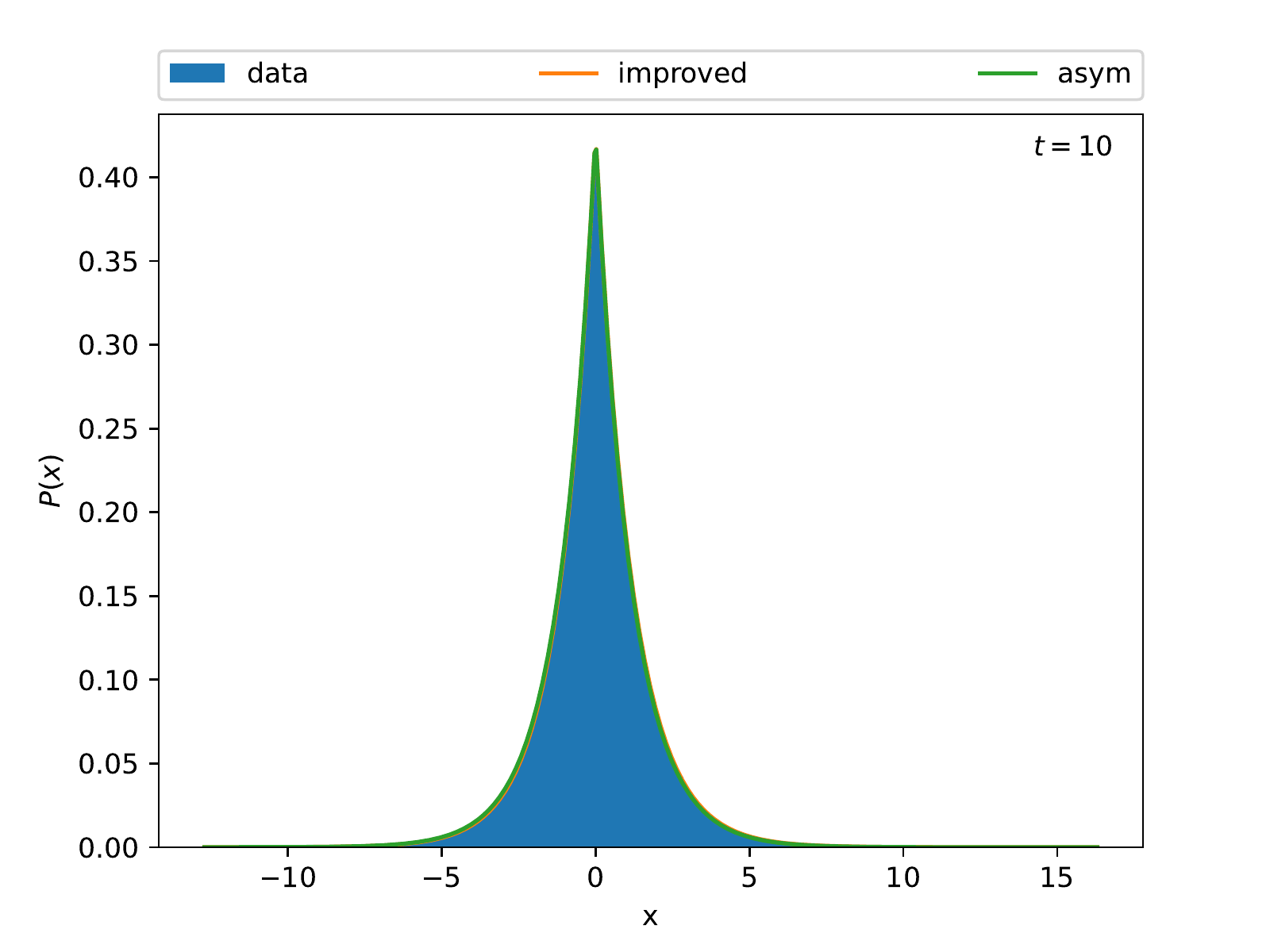}
	\newsubcap{(b)}
	\end{subfigure}
 	\begin{subfigure}{0.5\linewidth}
	\includegraphics[width=\columnwidth]{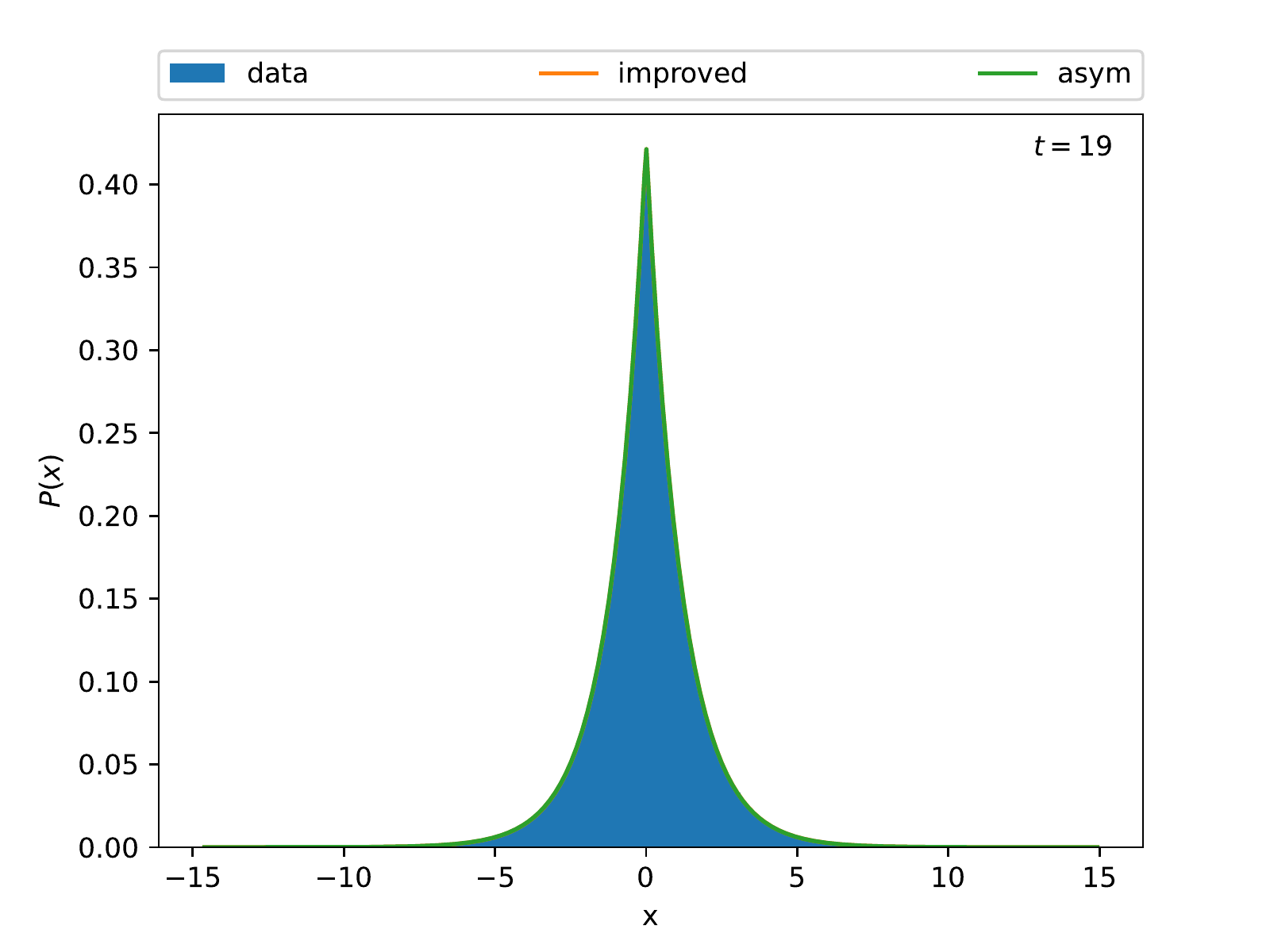}
	\newsubcap{(c) }
	\end{subfigure}
 \caption{Probability distribution of  $C_{\op}(t)$ for $s=0.52,\  L =40,\ \beta= 40,\ \caln = 10^6$ for $t=1$ in \capa for $t=10$ in \capb and for $t=19$ in \capc.
 }
  \label{Fig:$s=0.52,L=40$}
\end{figure}
\begin{figure}[!t]
 	\begin{subfigure}{0.5\linewidth}
    	\includegraphics[width=\columnwidth]{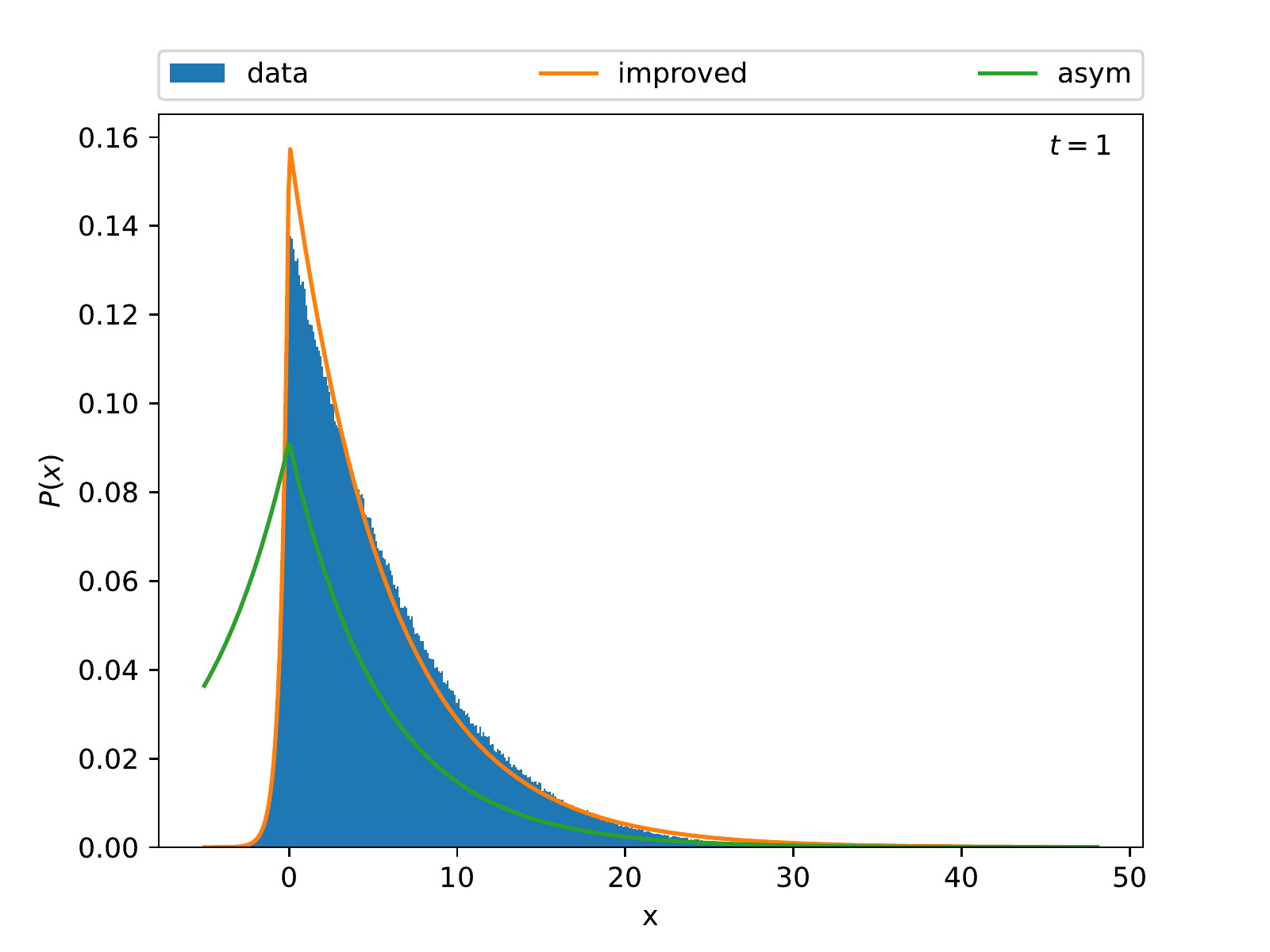}
    	\newsubcap{(a)}
	\end{subfigure}
	\begin{subfigure}{0.5\linewidth}
	\includegraphics[width=\columnwidth]{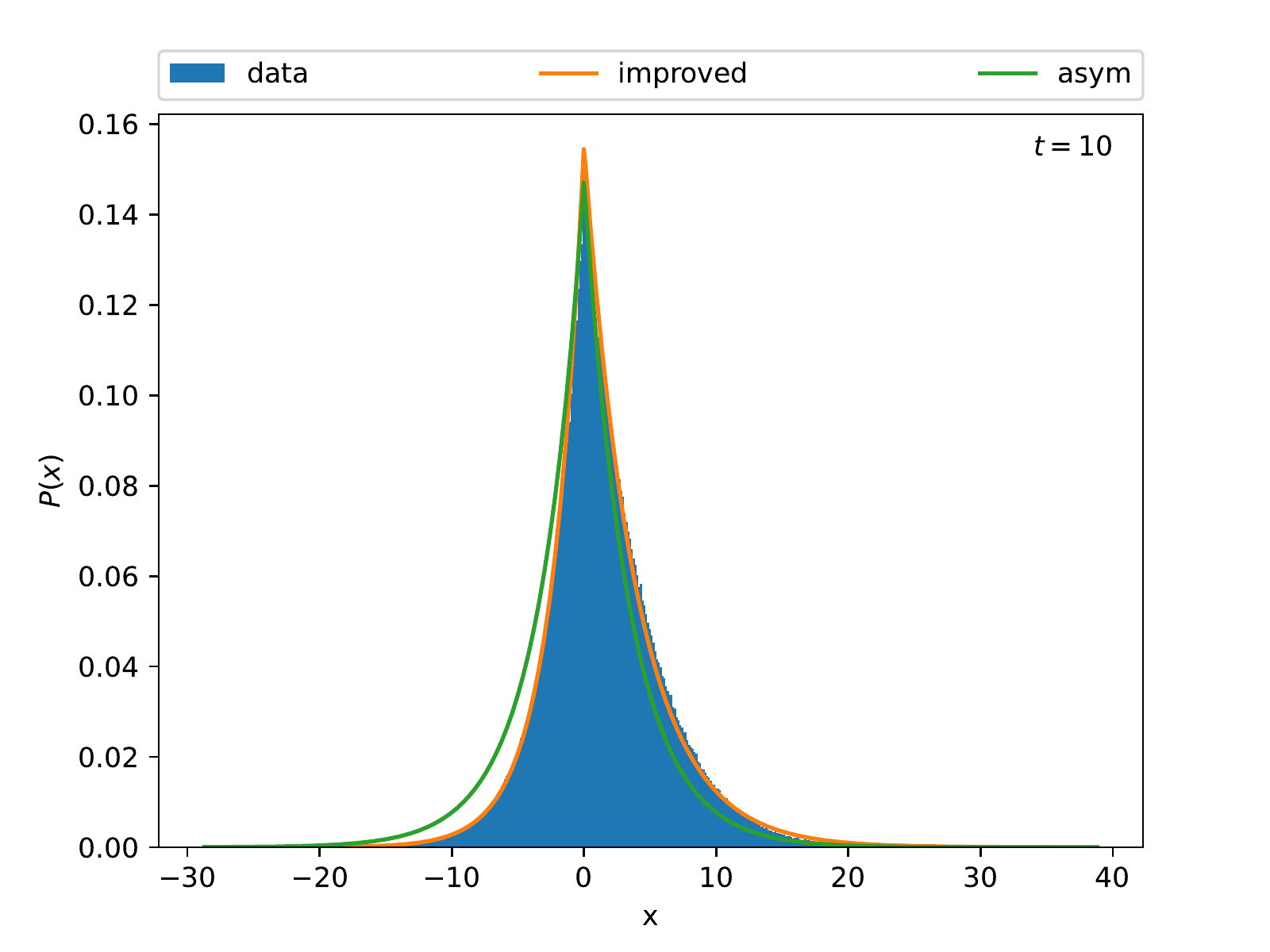}
	\newsubcap{(b)}
	\end{subfigure}
 	\begin{subfigure}{0.5\linewidth}
	\includegraphics[width=\columnwidth]{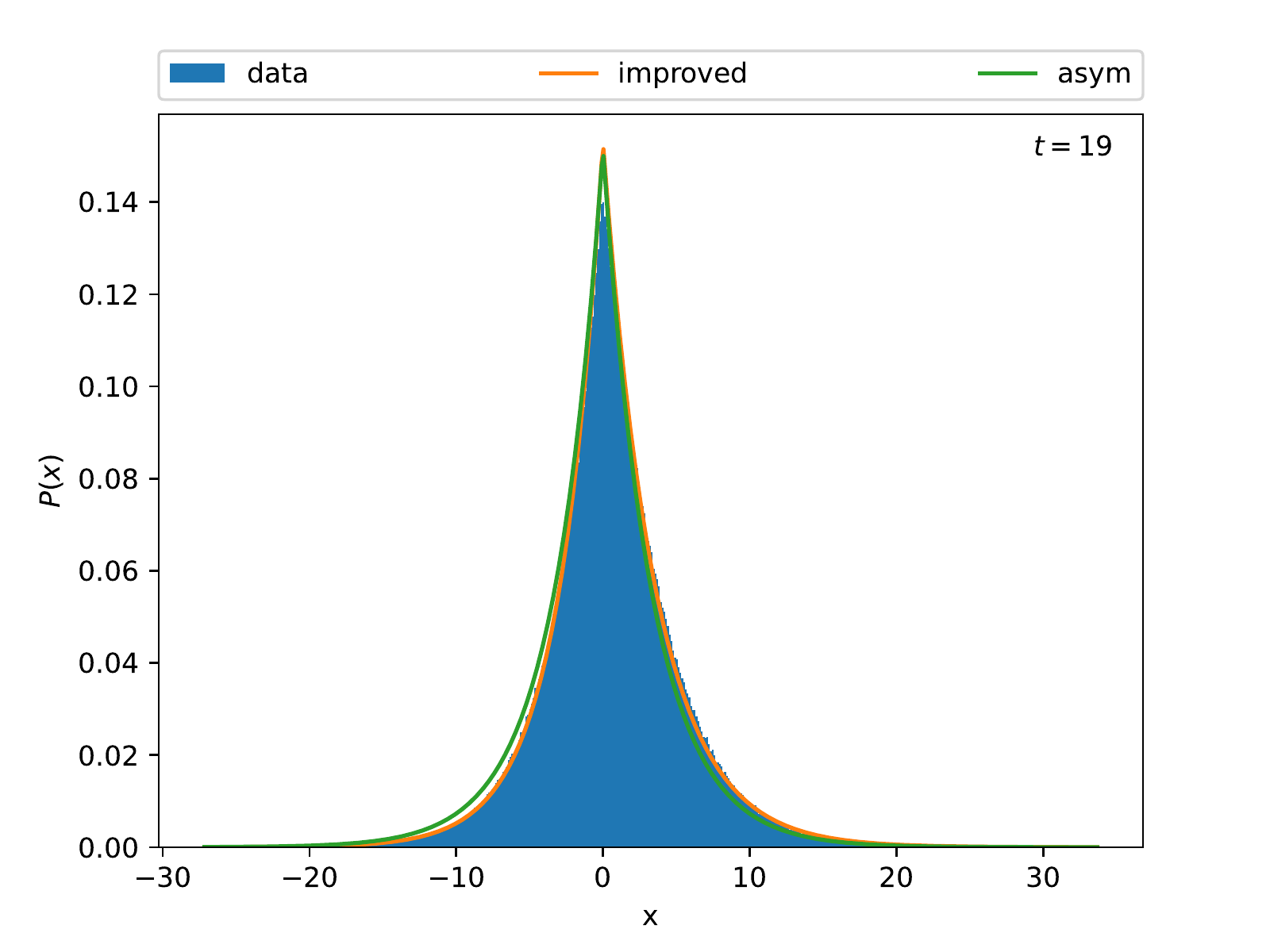}
	\newsubcap{(c) }
	\end{subfigure}
\caption{Probability distribution of  $C_{\op}(t)$ for $s=0.59,\ L =40,\ \beta= 40,\ \caln = 10^6$ for $t=1$ in \capa for $t=10$ in \capb and for $t=19$ in \capc.
}
\label{Fig:$s=0.59, L=40$}
\end{figure}
\begin{figure}[!t]
 	\begin{subfigure}{0.5\linewidth}
    	\includegraphics[width=\columnwidth]{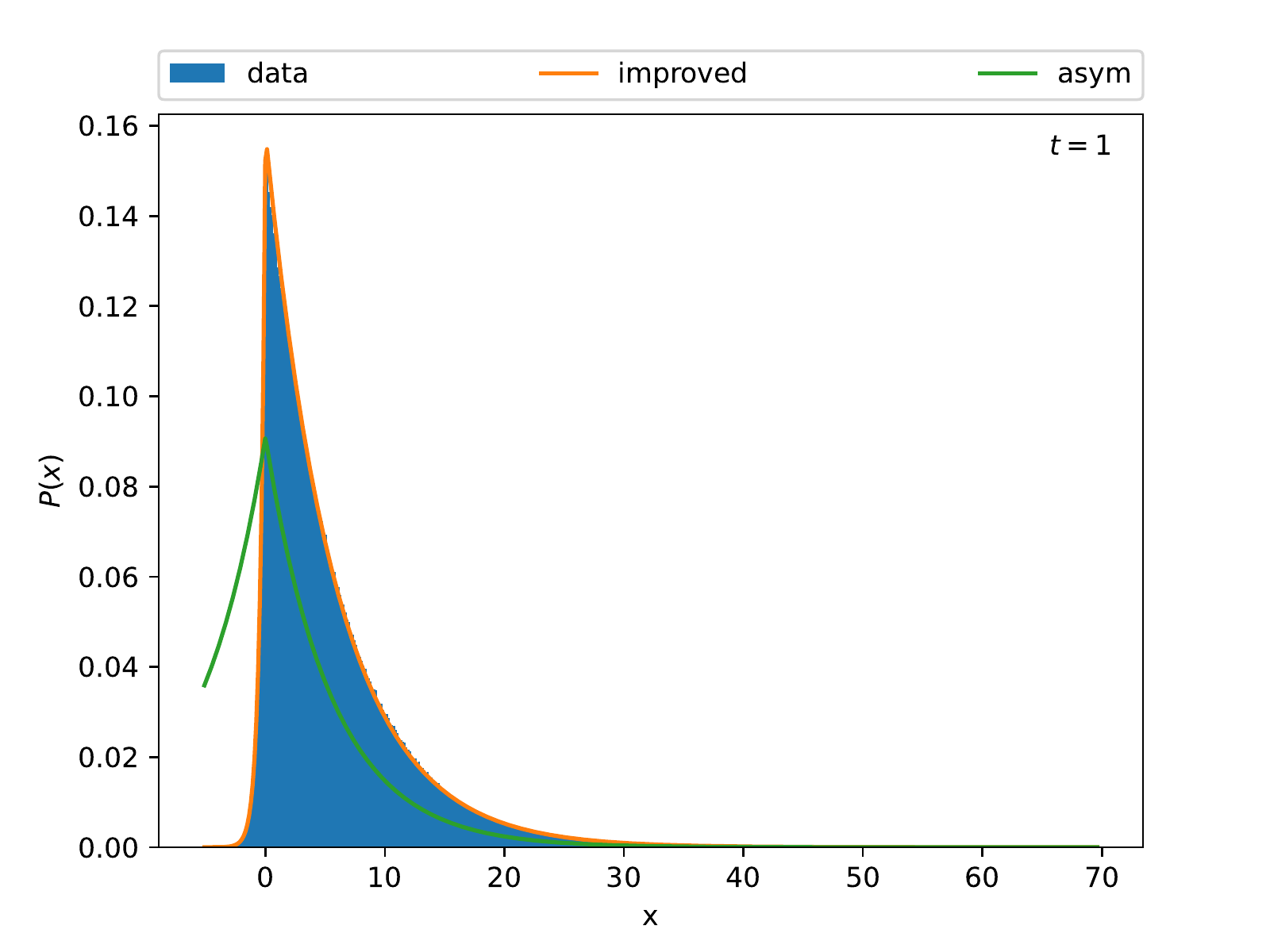}
    	\newsubcap{(a) }
	\end{subfigure}
	\begin{subfigure}{0.5\linewidth}
	\includegraphics[width=\columnwidth]{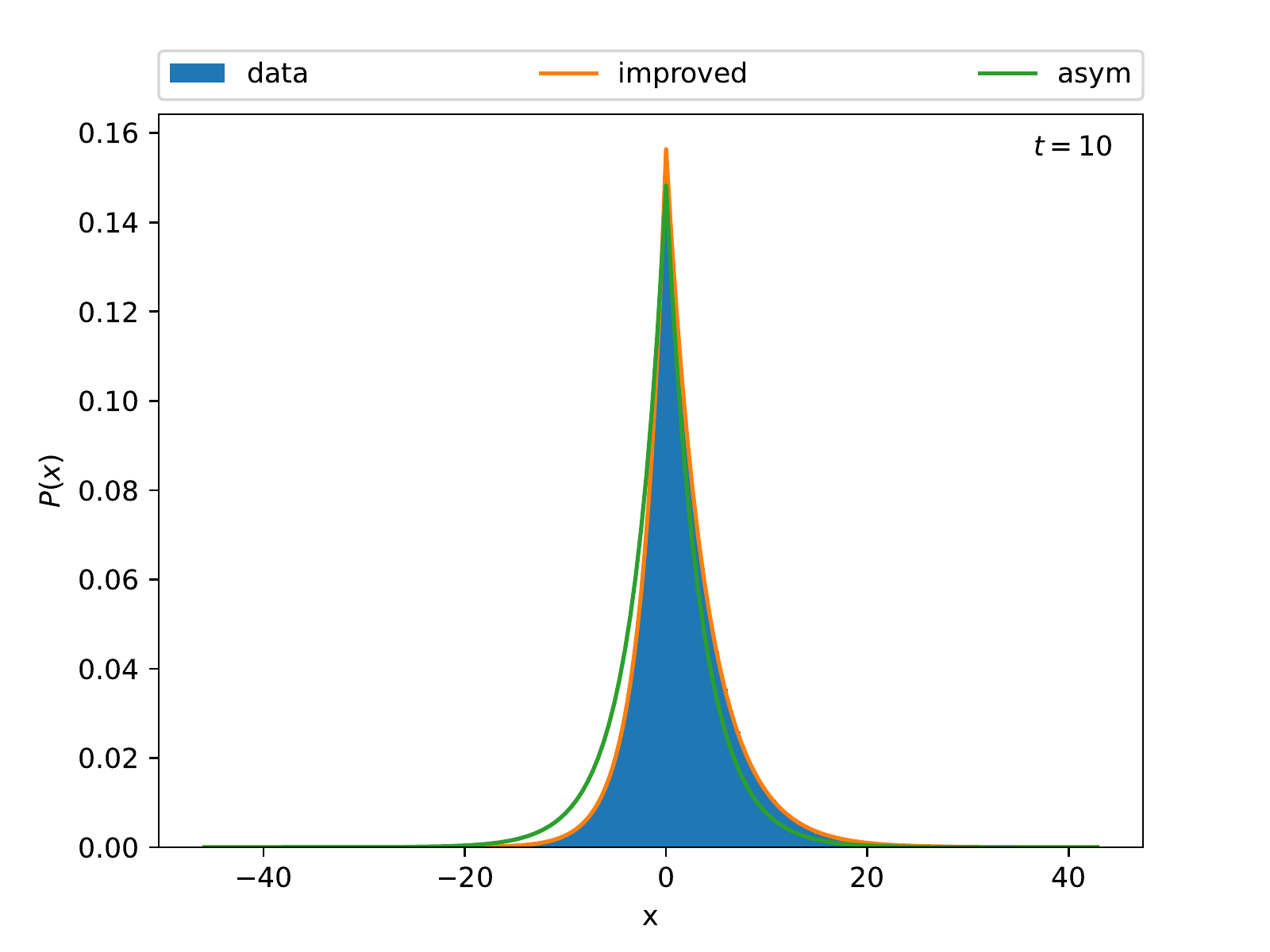}
	\newsubcap{(b) }
	\end{subfigure}
 	\begin{subfigure}{0.5\linewidth}
	\includegraphics[width=\columnwidth]{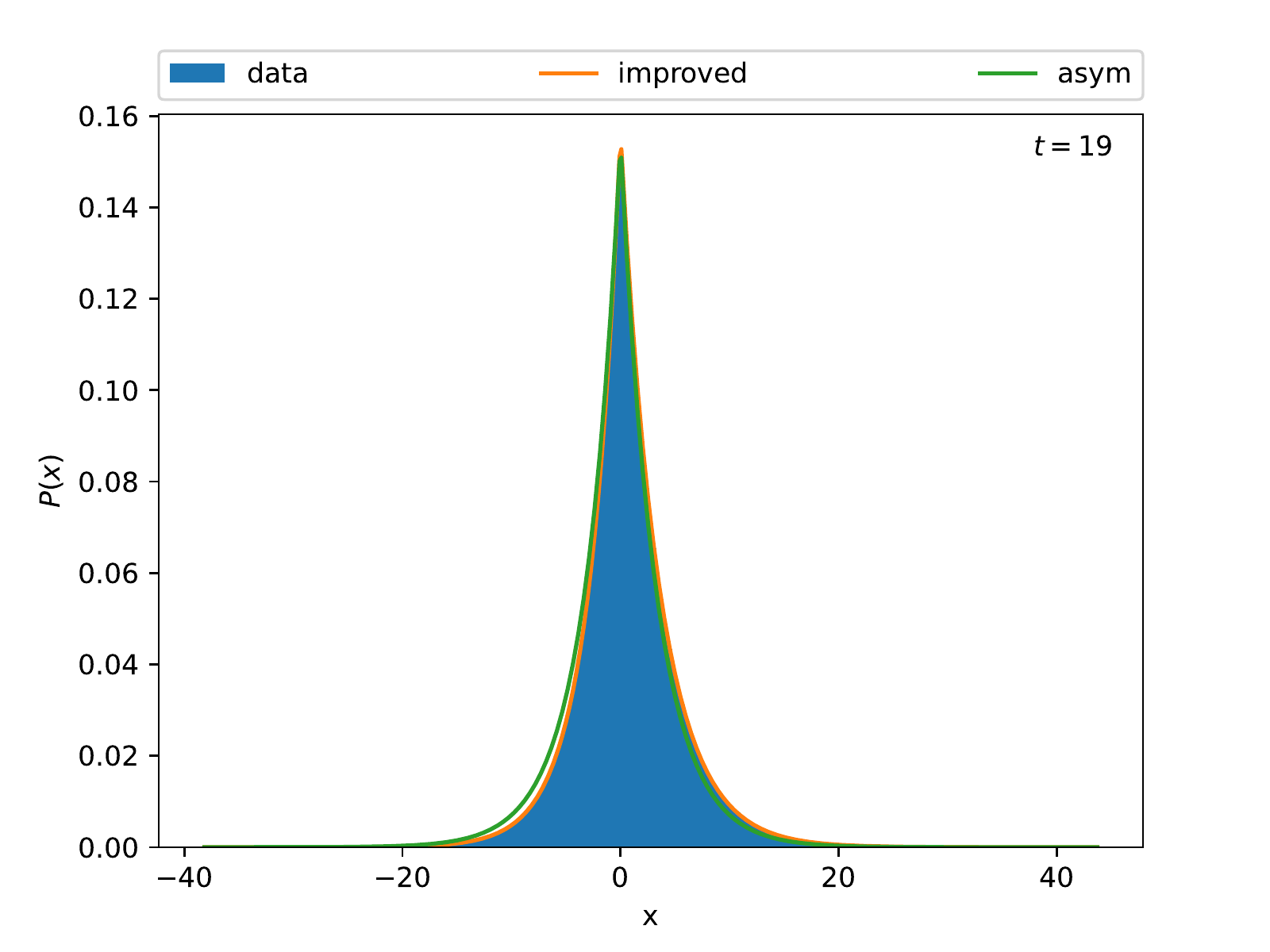}
	\newsubcap{(c) }
	\end{subfigure}
\caption{Probability distribution of  $C_{\op}(t)$ for $s=0.59,\ L =130,\ \beta= 40,\ \caln = 10^6$ for $t=1$ in \capa for $t=10$ in \capb and for $t=19$ in \capc.
}
\label{Fig:$s=0.59, L=130$}
\end{figure}
\begin{figure}[!t]
 	\begin{subfigure}{0.5\linewidth}
    	\includegraphics[width=\columnwidth]{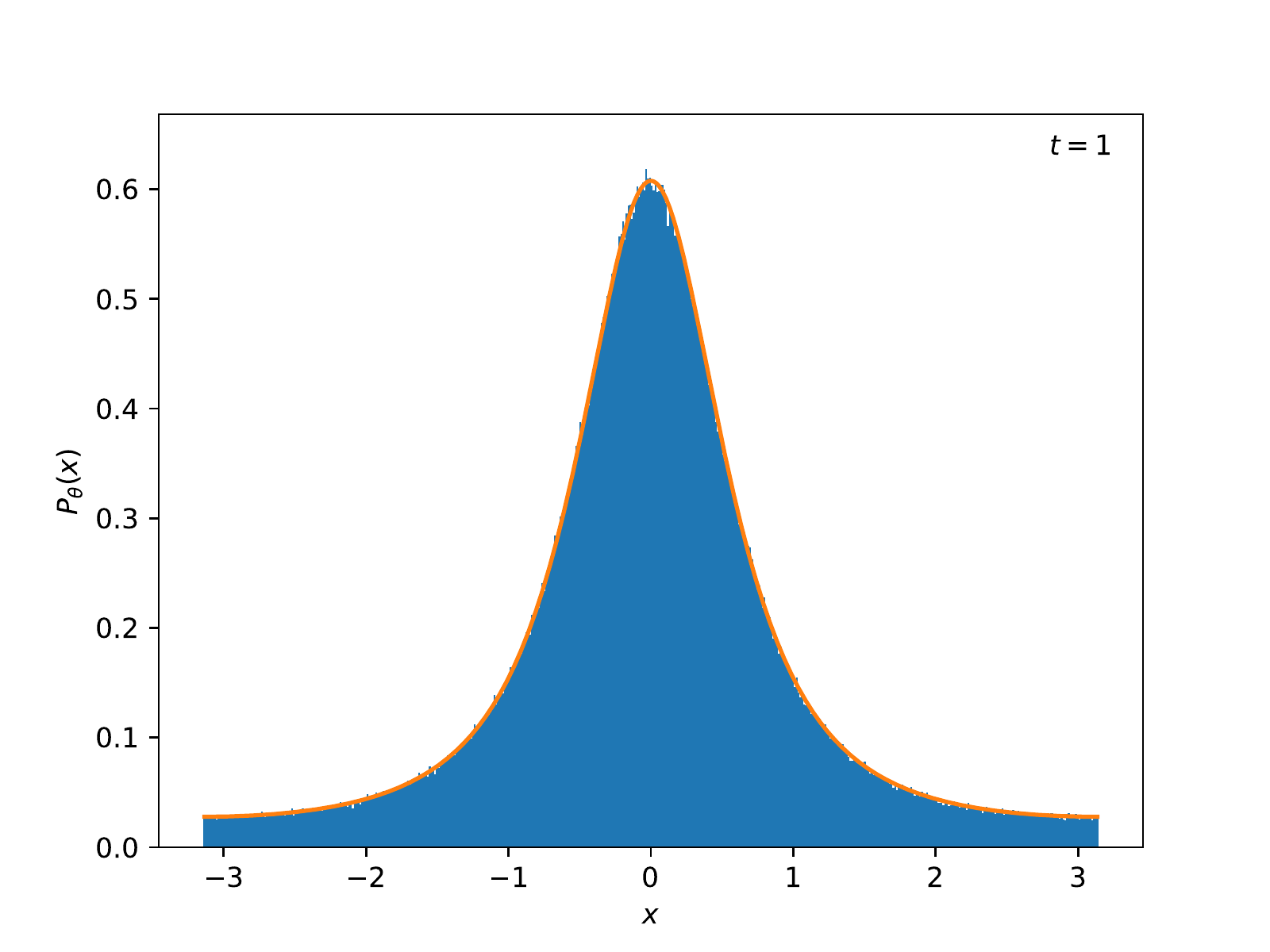}
    	\newsubcap{(a)}
	\end{subfigure}
	\begin{subfigure}{0.5\linewidth}
	\includegraphics[width=\columnwidth]{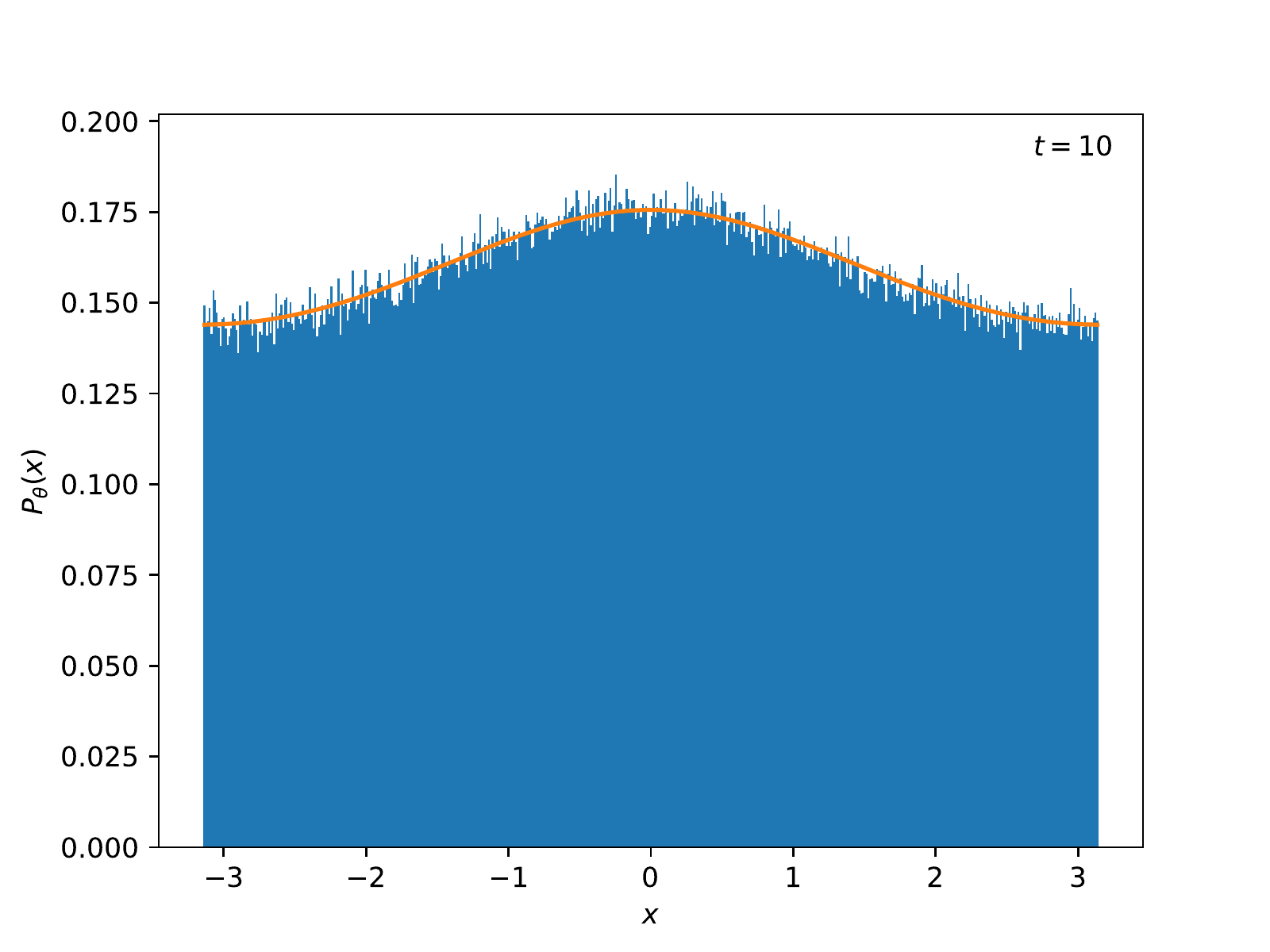}
	\newsubcap{(b)}
	\end{subfigure}
 	\begin{subfigure}{0.5\linewidth}
	\includegraphics[width=\columnwidth]{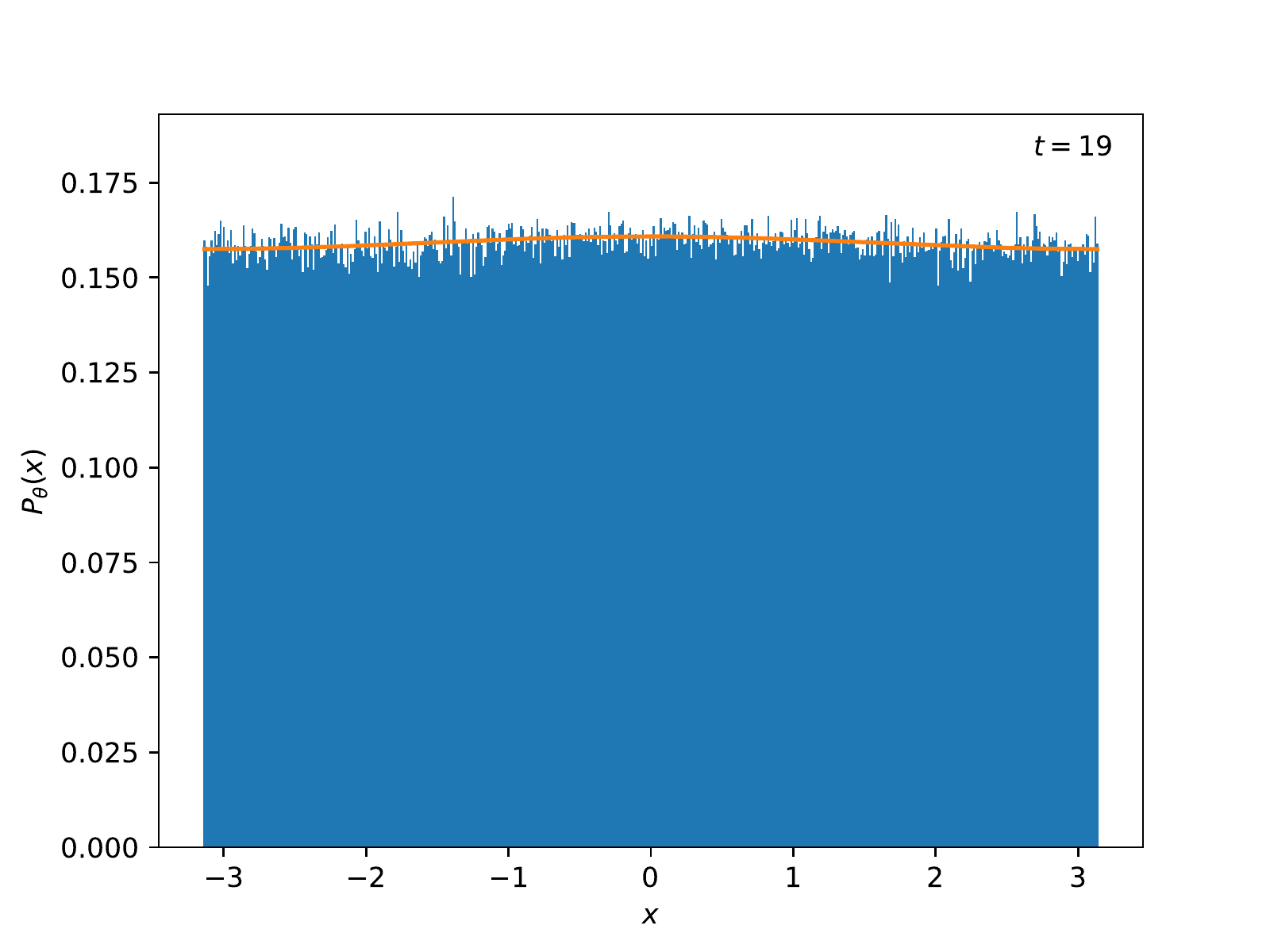}
	\newsubcap{(c)}

	\end{subfigure}
\caption{Probability distribution of the phase difference for $s=0.56,\ L =130,\ \beta= 40,\ \caln = 10^6$ for $t=1$ in \capa for $t=10$ in \capb and for $t=19$ in \capc.
}
\label{Fig:phase}
\end{figure}
\FloatBarrier

Finally, in Fig.\ \ref{Fig:phase}, histograms of the phase differences between the spatially-averaged fields at $0$ and $t$  are presented for various times for $s=0.56$.  As seen in previous studies \cite{Wagman:2016bam,Wagman:2017gqi,Wagman:2017xfh}, these phase distributions become increasingly broad as $t$ increases. The figure also shows fits of the asymptotic phase distribution given in Eq.\ \eqref{eq:phase_func} to these histograms; in all cases, this form provides an accurate description of the histograms, significantly improving on an assumption of wrapped-normality made in previous studies \cite{Detmold:2018eqd}.

\section{Improved estimators}
\label{sec:improved_estimator}
In this section, an improved estimator for the mean of $C_{Re}(t)$ will be presented. A detailed discussion of the concepts and examples considered in this section can be found in Ref.~\cite{Keener2010}. 
According to the Cram\'er-Rao bound, a classical result in estimation theory, if $\hat \theta$ is an unbiased estimator\footnote{In what follows, "hats" will be used to denote estimators, which depend on the given sample and therefore are random variables. } of a parameter $\t$, its variance $Var(\hat \t)$ satisfies the bound:
\bad 
Var(\hat \t ) \geq \ff{1}{\caln I(\t)},
\label{eq:Cramer-Rao}
\ead 
where $\caln$ is the sample size and $I(\t)$ is the Fisher information defined by:
\bad 
I(\t) = \int_{-\infty}^\ii dx\, P(x;\t)  \lp \frac{d \log P(x;\theta)}{d \t} \rp^2,
\ead 
where $P(x;\theta)$ is the distribution of the samples, parametrised by $\theta$. Asymptotically, when the maximum likelihood estimator exists, it saturates the Cram\'er-Rao bound and therefore has the least variance among all unbiased estimators. However, in some cases there are other examples of minimum-variance unbiased estimators for finite sample size. As a simple example\footnote{In this case, the Fisher information $I(\t)$ is divergent and therefore the Cram\'er-Rao bound is not available.}, if one considers the uniform distribution supported on the interval $[0,\t]$, the minimum-variance unbiased estimator of the mean of $\t$ is given by:
\bad 
\hat \m_{U,mvue} = \ff{\caln+1}{2\caln} \max_i x_i.
\ead 
In fact, for this example, the ratio of the variance of the minimum-variance unbiased estimator defined above to the variance of the sample mean $\hat \mu_{U,sm}$ can be calculated easily and is found to be:
\begin{equation}
\ff{Var \lp \hat \mu_{U,mvue}\rp}{Var \lp \hat \mu_{U,sm} \rp} = \ff{3}{\caln+2},
\end{equation}
which vanishes as the sample size goes to infinity. In certain cases, an estimator with smaller mean squared error\footnote{MSE of an estimator $\hat \mu$ of $\mu$ is defined by $MSE(\hat \mu) = \ev{\lp \hat \mu - \mu \rp^2}$.} (MSE) can be constructed if a bias is accepted. For example, for the example considered above, a biased estimator $\hat \mu_{U,b}$ can be constructed as
\bad 
\hat \mu_{U,b} = \ff{\caln+2}{2(\caln+1)} \max_i x_i,
\ead 
whose MSE is less than that of $\hat \m_{U,mvue}$. 

In the context of the $O(2)$ model correlation functions, using the analytic form of the PDF in Eq.~\eqref{eq:C_2d_dists} for the real part of the correlation function, it can be shown that the sample mean $\hat \mu_{Re,sm}$ is efficient within the class of unbiased estimators of $\mu_{Re}$. According to the bivariate generalization of the Cram\'er-Rao bound, given the parameters $\vec\t = \{\t_1,\t_2\}$, the variance of the an unbiased estimator $\hat T$ of a quantity $T = T(\t_1,\t_2)$ is given by:
\bad 
Var(\hat T) \geq \ff{1}{\caln} \sum_{i,j=1}^2\pdv{T}{\t_i} \pdv{T}{\t_j} I^{-1}_{ij},
\label{eq:Cramer-Rao-multi}
\ead 
where $I_{ij}$ is defined by:
\bad 
I_{ij} = \int_{-\ii}^\ii dx\, P(x;\vec\t) \pdv{\log P(x;\vec\t)}{\t_i}\pdv{\log P(x;\vec\t)}{\t_j}.
\ead 
It is convenient to choose the parameters as $\t_1 = \o_{Re,+}$, $\t_2 = \o_{Re,-}$. From Eq. \eqref{eq:C_2d_dists}, it follows that:
\bad 
\mu_{Re} &= \ff{1}{\o_{R,+}}-\ff{1}{\o_{R,-}},\\ 
\s^2_{Re} &= \ff{1}{\o_{R,+}^2}+\ff{1}{\o_{R,-}^2}.
\label{eq:Re_eqs}
\ead
Choosing $T = \mu_{Re}$, it is found from Eqs. \eqref{eq:Cramer-Rao-multi} and \eqref{eq:Re_eqs} that $Var(\hat T) \geq \ff{1}{\caln}\s^2_{Re}$ for all unbiased estimators $\hat T$ of $T$. For the sample mean $\hat \mu_{Re,sm} = \ff{1}{\caln}\sum_{i=1}^\caln x_i$, the variance is $Var(\hat \mu_{Re,sm}) = \ff{1}{\caln}\s^2_{Re}$, so  the sample mean  is the minimum variance unbiased estimator in this case.

The PDF of $C_{Re}(t)$ given in Eq.~\eqref{eq:C_2d_dists} also allows for the construction of improved estimators that outperform the sample mean of $C_{Re}(t)$.
In particular, a biased   estimator for $\mu_{Re}$ can be defined as:
\bad 
\hat \mu_{Re,b} = \ff{1}{\caln +1}\sum_{i=1}^\caln x_{i}.
\ead 
The mean squared error for $\hat \mu_{Re,b}$ is given by:
\begin{widetext}
\bad 
MSE(\hat \mu_{Re,b}) &= \ev{\lp \hat \mu_{Re,b}-\mu_{Re} \rp ^2 } \\ 
&= \ev{\ff{1}{(\caln+1)^2} \lp \sum_{i=1}^\caln x_i \rp^2 - \ff{2}{\caln + 1}\mu_{Re}\sum_{i=1}^\caln x_i + \mu_{Re}^2} \\ 
&= \ff{1}{(\caln+1)^2} \lp \caln \lp \s_{Re}^2 + \mu_{Re}^2 \rp + \caln \lp \caln-1\rp \mu_{Re}^2 \rp - \ff{\caln - 1}{\caln +1}\mu_{Re}^2 \\ 
&= \ff{\caln \s_{Re}^2 + \mu_{Re}^2 }{\lp \caln +1 \rp^2}.
\ead 
\label{eq:MSE_mu_Re_b}
\end{widetext}
From Eq. \eqref{eq:Re_eqs} and the form of the PDF, it follows that $\s_{Re}^2 > \mu_{Re}^2$. Therefore, $MSE(\hat \mu_{Re,b}) < \ff{\s_{Re}^2}{\caln + 1}$ and
\bad 
\ff{MSE(\hat \mu_{Re,b})}{MSE(\hat \mu_{Re,sm})} < \ff{\caln}{\caln + 1},
\ead 
showing that $\hat \mu_{Re,b}$ is a marginally more efficient estimator of the mean than the sample mean, albeit a biased one. The bias of $\hat \mu_{Re,b}$ can be calculated easily and is found to be:
\bad
Bias(\hat \mu_{Re,b}) = -\ff{\mu_{Re}}{\caln + 1},
\ead 
which vanishes as $\caln \to \ii$.
While $\hat\mu_{Re,b}$ is only marginally more efficient than the sample mean  for ${\cal N}\to\infty$, other biased estimators may exist that improve on this behaviour. With recent interest in so-called ``master field'' Monte-Carlo calculations \cite{Luscher:2017cjh}, where a small number of samples of large volume lattice geometries are envisioned, even non-asymptotic improvements are of interest.

\section{Conclusions}
\label{sec:conc}
In the present work, the exact probability distributions of the two-point correlation functions of the $O(N)$ model in the disordered phase at vanishing spatial momentum and infinite volume have been derived. Numerical tests have been performed for the $N=2$ case and are found to support the assumptions needed in the analytic derivations. While the theories discussed in the present work are relatively simple, the insights gained from having an analytic description of the PDF of correlation functions are quite far-reaching. They provide a means of assessing the reliability of numerical Monte-Carlo calculations and an alternative method whereby the spectrum of a theory can be accessed from a single timeslice \cite{Yunus:2022pto}. As shown in Sec. \ref{sec:improved_estimator} for the case of the $O(2)$ model, knowledge of the PDF allows better estimators of the expectation value of a correlation function than the sample mean. If such estimators could be found for the expectation values of correlation functions in lattice quantum field theories in general, it would provide an interesting path towards addressing the commonly-occurring signal-to-noise problem. 

As discussed earlier, the arguments presented here generalize readily to correlation functions of bosonic theories without an $O(N)$ symmetry although the expressions for the PDFs become more complicated. For $N=1$, the discussion above can be\footnote{Except when $d=1$. In this case, solitonic vacua also appear and as these vacua do not have a mass gap, further discussion is needed.}  extended to the case of the broken phase by expressing the probability distributions as a superposition of correlation function distributions associated with the multiple different vacua. For $N>1$, the broken phase has Goldstone bosons and therefore has infinite correlation length in the absence of explicit symmetry-breaking. Consequently, the arguments presented here do not immediately generalize to this case. 
In principle, the arguments used in this study can be applied to higher-point correlation functions, although the resulting expressions for PDFs are expected to be more complicated.
Finally, it is likely that the framework developed here can be adapted to correlation functions of glueball interpolating operators and of large Wilson loops in pure gauge theories. These directions will be explored in subsequent work.

\acknowledgements{ 
We are grateful to M. L. Wagman for discussions.
This work is supported by the National Science Foundation under Cooperative Agreement PHY-2019786 (The NSF AI Institute for Artificial Intelligence and Fundamental Interactions, http://iaifi.org/) and by the U.S.~Department of Energy, Office of Science, Office of Nuclear Physics under contract number DE-SC0011090. WD is also supported by the SciDAC5 award DE-SC0023116. 
}

\bibliography{refs}

\begin{thebibliography}{31}%
\makeatletter
\providecommand \@ifxundefined [1]{%
 \@ifx{#1\undefined}
}%
\providecommand \@ifnum [1]{%
 \ifnum #1\expandafter \@firstoftwo
 \else \expandafter \@secondoftwo
 \fi
}%
\providecommand \@ifx [1]{%
 \ifx #1\expandafter \@firstoftwo
 \else \expandafter \@secondoftwo
 \fi
}%
\providecommand \natexlab [1]{#1}%
\providecommand \enquote  [1]{``#1''}%
\providecommand \bibnamefont  [1]{#1}%
\providecommand \bibfnamefont [1]{#1}%
\providecommand \citenamefont [1]{#1}%
\providecommand \href@noop [0]{\@secondoftwo}%
\providecommand \href [0]{\begingroup \@sanitize@url \@href}%
\providecommand \@href[1]{\@@startlink{#1}\@@href}%
\providecommand \@@href[1]{\endgroup#1\@@endlink}%
\providecommand \@sanitize@url [0]{\catcode `\\12\catcode `\$12\catcode
  `\&12\catcode `\#12\catcode `\^12\catcode `\_12\catcode `\%12\relax}%
\providecommand \@@startlink[1]{}%
\providecommand \@@endlink[0]{}%
\providecommand \url  [0]{\begingroup\@sanitize@url \@url }%
\providecommand \@url [1]{\endgroup\@href {#1}{\urlprefix }}%
\providecommand \urlprefix  [0]{URL }%
\providecommand \Eprint [0]{\href }%
\providecommand \doibase [0]{http://dx.doi.org/}%
\providecommand \selectlanguage [0]{\@gobble}%
\providecommand \bibinfo  [0]{\@secondoftwo}%
\providecommand \bibfield  [0]{\@secondoftwo}%
\providecommand \translation [1]{[#1]}%
\providecommand \BibitemOpen [0]{}%
\providecommand \bibitemStop [0]{}%
\providecommand \bibitemNoStop [0]{.\EOS\space}%
\providecommand \EOS [0]{\spacefactor3000\relax}%
\providecommand \BibitemShut  [1]{\csname bibitem#1\endcsname}%
\let\auto@bib@innerbib\@empty
%</preamble>
\bibitem [{\citenamefont {Della~Morte}\ and\ \citenamefont
  {Giusti}(2009{\natexlab{a}})}]{DellaMorte:2007zz}%
  \BibitemOpen
  \bibfield  {author} {\bibinfo {author} {\bibfnamefont {M.}~\bibnamefont
  {Della~Morte}}\ and\ \bibinfo {author} {\bibfnamefont {L.}~\bibnamefont
  {Giusti}},\ }\href {\doibase 10.1016/j.cpc.2008.10.017} {\bibfield  {journal}
  {\bibinfo  {journal} {Comput. Phys. Commun.}\ }\textbf {\bibinfo {volume}
  {180}},\ \bibinfo {pages} {813} (\bibinfo {year}
  {2009}{\natexlab{a}})}\BibitemShut {NoStop}%
\bibitem [{\citenamefont {Della~Morte}\ and\ \citenamefont
  {Giusti}(2009{\natexlab{b}})}]{DellaMorte:2008jd}%
  \BibitemOpen
  \bibfield  {author} {\bibinfo {author} {\bibfnamefont {M.}~\bibnamefont
  {Della~Morte}}\ and\ \bibinfo {author} {\bibfnamefont {L.}~\bibnamefont
  {Giusti}},\ }\href {\doibase 10.1016/j.cpc.2009.03.009} {\bibfield  {journal}
  {\bibinfo  {journal} {Comput. Phys. Commun.}\ }\textbf {\bibinfo {volume}
  {180}},\ \bibinfo {pages} {819} (\bibinfo {year} {2009}{\natexlab{b}})},\
  \Eprint {http://arxiv.org/abs/0806.2601} {arXiv:0806.2601 [hep-lat]}
  \BibitemShut {NoStop}%
\bibitem [{\citenamefont {Della~Morte}\ and\ \citenamefont
  {Giusti}(2011)}]{DellaMorte:2010yp}%
  \BibitemOpen
  \bibfield  {author} {\bibinfo {author} {\bibfnamefont {M.}~\bibnamefont
  {Della~Morte}}\ and\ \bibinfo {author} {\bibfnamefont {L.}~\bibnamefont
  {Giusti}},\ }\href {\doibase 10.1007/JHEP05(2011)056} {\bibfield  {journal}
  {\bibinfo  {journal} {JHEP}\ }\textbf {\bibinfo {volume} {05}},\ \bibinfo
  {pages} {056} (\bibinfo {year} {2011})},\ \Eprint
  {http://arxiv.org/abs/1012.2562} {arXiv:1012.2562 [hep-lat]} \BibitemShut
  {NoStop}%
\bibitem [{\citenamefont {Detmold}\ and\ \citenamefont
  {Endres}(2014)}]{Detmold:2014hla}%
  \BibitemOpen
  \bibfield  {author} {\bibinfo {author} {\bibfnamefont {W.}~\bibnamefont
  {Detmold}}\ and\ \bibinfo {author} {\bibfnamefont {M.~G.}\ \bibnamefont
  {Endres}},\ }\href {\doibase 10.1103/PhysRevD.90.034503} {\bibfield
  {journal} {\bibinfo  {journal} {Phys. Rev. D}\ }\textbf {\bibinfo {volume}
  {90}},\ \bibinfo {pages} {034503} (\bibinfo {year} {2014})},\ \Eprint
  {http://arxiv.org/abs/1404.6816} {arXiv:1404.6816 [hep-lat]} \BibitemShut
  {NoStop}%
\bibitem [{\citenamefont {Majumdar}\ \emph {et~al.}(2014)\citenamefont
  {Majumdar}, \citenamefont {Mathur},\ and\ \citenamefont
  {Mondal}}]{Majumdar:2014cqa}%
  \BibitemOpen
  \bibfield  {author} {\bibinfo {author} {\bibfnamefont {P.}~\bibnamefont
  {Majumdar}}, \bibinfo {author} {\bibfnamefont {N.}~\bibnamefont {Mathur}}, \
  and\ \bibinfo {author} {\bibfnamefont {S.}~\bibnamefont {Mondal}},\ }\href
  {\doibase 10.1016/j.physletb.2014.07.056} {\bibfield  {journal} {\bibinfo
  {journal} {Phys. Lett. B}\ }\textbf {\bibinfo {volume} {736}},\ \bibinfo
  {pages} {415} (\bibinfo {year} {2014})},\ \Eprint
  {http://arxiv.org/abs/1403.2936} {arXiv:1403.2936 [hep-lat]} \BibitemShut
  {NoStop}%
\bibitem [{\citenamefont {C\`e}\ \emph {et~al.}(2017)\citenamefont {C\`e},
  \citenamefont {Giusti},\ and\ \citenamefont {Schaefer}}]{Ce:2016ajy}%
  \BibitemOpen
  \bibfield  {author} {\bibinfo {author} {\bibfnamefont {M.}~\bibnamefont
  {C\`e}}, \bibinfo {author} {\bibfnamefont {L.}~\bibnamefont {Giusti}}, \ and\
  \bibinfo {author} {\bibfnamefont {S.}~\bibnamefont {Schaefer}},\ }\href
  {\doibase 10.1103/PhysRevD.95.034503} {\bibfield  {journal} {\bibinfo
  {journal} {Phys. Rev. D}\ }\textbf {\bibinfo {volume} {95}},\ \bibinfo
  {pages} {034503} (\bibinfo {year} {2017})},\ \Eprint
  {http://arxiv.org/abs/1609.02419} {arXiv:1609.02419 [hep-lat]} \BibitemShut
  {NoStop}%
\bibitem [{\citenamefont {C\`e}\ \emph {et~al.}(2016)\citenamefont {C\`e},
  \citenamefont {Giusti},\ and\ \citenamefont {Schaefer}}]{Ce:2016idq}%
  \BibitemOpen
  \bibfield  {author} {\bibinfo {author} {\bibfnamefont {M.}~\bibnamefont
  {C\`e}}, \bibinfo {author} {\bibfnamefont {L.}~\bibnamefont {Giusti}}, \ and\
  \bibinfo {author} {\bibfnamefont {S.}~\bibnamefont {Schaefer}},\ }\href
  {\doibase 10.1103/PhysRevD.93.094507} {\bibfield  {journal} {\bibinfo
  {journal} {Phys. Rev. D}\ }\textbf {\bibinfo {volume} {93}},\ \bibinfo
  {pages} {094507} (\bibinfo {year} {2016})},\ \Eprint
  {http://arxiv.org/abs/1601.04587} {arXiv:1601.04587 [hep-lat]} \BibitemShut
  {NoStop}%
\bibitem [{\citenamefont {Wagman}\ and\ \citenamefont
  {Savage}(2017{\natexlab{a}})}]{Wagman:2017xfh}%
  \BibitemOpen
  \bibfield  {author} {\bibinfo {author} {\bibfnamefont {M.~L.}\ \bibnamefont
  {Wagman}}\ and\ \bibinfo {author} {\bibfnamefont {M.~J.}\ \bibnamefont
  {Savage}},\ }\href@noop {} {\  (\bibinfo {year} {2017}{\natexlab{a}})},\
  \Eprint {http://arxiv.org/abs/1704.07356} {arXiv:1704.07356 [hep-lat]}
  \BibitemShut {NoStop}%
\bibitem [{\citenamefont {Wagman}(2017)}]{Wagman:2017gqi}%
  \BibitemOpen
  \bibfield  {author} {\bibinfo {author} {\bibfnamefont {M.~L.}\ \bibnamefont
  {Wagman}},\ }\emph {\bibinfo {title} {{Statistical Angles on the Lattice QCD
  Signal-to-Noise Problem}}},\ \href@noop {} {Ph.D. thesis},\ \bibinfo
  {school} {U. Washington, Seattle (main)} (\bibinfo {year} {2017}),\ \Eprint
  {http://arxiv.org/abs/1711.00062} {arXiv:1711.00062 [hep-lat]} \BibitemShut
  {NoStop}%
\bibitem [{\citenamefont {Detmold}\ \emph {et~al.}(2018)\citenamefont
  {Detmold}, \citenamefont {Kanwar},\ and\ \citenamefont
  {Wagman}}]{Detmold:2018eqd}%
  \BibitemOpen
  \bibfield  {author} {\bibinfo {author} {\bibfnamefont {W.}~\bibnamefont
  {Detmold}}, \bibinfo {author} {\bibfnamefont {G.}~\bibnamefont {Kanwar}}, \
  and\ \bibinfo {author} {\bibfnamefont {M.~L.}\ \bibnamefont {Wagman}},\
  }\href {\doibase 10.1103/PhysRevD.98.074511} {\bibfield  {journal} {\bibinfo
  {journal} {Phys. Rev. D}\ }\textbf {\bibinfo {volume} {98}},\ \bibinfo
  {pages} {074511} (\bibinfo {year} {2018})},\ \Eprint
  {http://arxiv.org/abs/1806.01832} {arXiv:1806.01832 [hep-lat]} \BibitemShut
  {NoStop}%
\bibitem [{\citenamefont {Porter}\ and\ \citenamefont
  {Drut}(2017)}]{Porter:2016vry}%
  \BibitemOpen
  \bibfield  {author} {\bibinfo {author} {\bibfnamefont {W.~J.}\ \bibnamefont
  {Porter}}\ and\ \bibinfo {author} {\bibfnamefont {J.~E.}\ \bibnamefont
  {Drut}},\ }\href {\doibase 10.1103/PhysRevA.95.053619} {\bibfield  {journal}
  {\bibinfo  {journal} {Phys. Rev. A}\ }\textbf {\bibinfo {volume} {95}},\
  \bibinfo {pages} {053619} (\bibinfo {year} {2017})},\ \Eprint
  {http://arxiv.org/abs/1609.09401} {arXiv:1609.09401 [cond-mat.quant-gas]}
  \BibitemShut {NoStop}%
\bibitem [{\citenamefont {Dalla~Brida}\ \emph {et~al.}(2021)\citenamefont
  {Dalla~Brida}, \citenamefont {Giusti}, \citenamefont {Harris},\ and\
  \citenamefont {Pepe}}]{DallaBrida:2020cik}%
  \BibitemOpen
  \bibfield  {author} {\bibinfo {author} {\bibfnamefont {M.}~\bibnamefont
  {Dalla~Brida}}, \bibinfo {author} {\bibfnamefont {L.}~\bibnamefont {Giusti}},
  \bibinfo {author} {\bibfnamefont {T.}~\bibnamefont {Harris}}, \ and\ \bibinfo
  {author} {\bibfnamefont {M.}~\bibnamefont {Pepe}},\ }\href {\doibase
  10.1016/j.physletb.2021.136191} {\bibfield  {journal} {\bibinfo  {journal}
  {Phys. Lett. B}\ }\textbf {\bibinfo {volume} {816}},\ \bibinfo {pages}
  {136191} (\bibinfo {year} {2021})},\ \Eprint
  {http://arxiv.org/abs/2007.02973} {arXiv:2007.02973 [hep-lat]} \BibitemShut
  {NoStop}%
\bibitem [{\citenamefont {Detmold}\ \emph {et~al.}(2020)\citenamefont
  {Detmold}, \citenamefont {Kanwar}, \citenamefont {Wagman},\ and\
  \citenamefont {Warrington}}]{Detmold:2020ncp}%
  \BibitemOpen
  \bibfield  {author} {\bibinfo {author} {\bibfnamefont {W.}~\bibnamefont
  {Detmold}}, \bibinfo {author} {\bibfnamefont {G.}~\bibnamefont {Kanwar}},
  \bibinfo {author} {\bibfnamefont {M.~L.}\ \bibnamefont {Wagman}}, \ and\
  \bibinfo {author} {\bibfnamefont {N.~C.}\ \bibnamefont {Warrington}},\ }\href
  {\doibase 10.1103/PhysRevD.102.014514} {\bibfield  {journal} {\bibinfo
  {journal} {Phys. Rev. D}\ }\textbf {\bibinfo {volume} {102}},\ \bibinfo
  {pages} {014514} (\bibinfo {year} {2020})},\ \Eprint
  {http://arxiv.org/abs/2003.05914} {arXiv:2003.05914 [hep-lat]} \BibitemShut
  {NoStop}%
\bibitem [{\citenamefont {Kanwar}(2021)}]{Kanwar:2021wzm}%
  \BibitemOpen
  \bibfield  {author} {\bibinfo {author} {\bibfnamefont {G.}~\bibnamefont
  {Kanwar}},\ }\emph {\bibinfo {title} {{Machine Learning and Variational
  Algorithms for Lattice Field Theory}}},\ \href@noop {} {\bibinfo {type}
  {Ph.d. thesis}} (\bibinfo {year} {2021}),\ \Eprint
  {http://arxiv.org/abs/2106.01975} {arXiv:2106.01975 [hep-lat]} \BibitemShut
  {NoStop}%
\bibitem [{\citenamefont {Detmold}\ \emph {et~al.}(2021)\citenamefont
  {Detmold}, \citenamefont {Kanwar}, \citenamefont {Lamm}, \citenamefont
  {Wagman},\ and\ \citenamefont {Warrington}}]{Detmold:2021ulb}%
  \BibitemOpen
  \bibfield  {author} {\bibinfo {author} {\bibfnamefont {W.}~\bibnamefont
  {Detmold}}, \bibinfo {author} {\bibfnamefont {G.}~\bibnamefont {Kanwar}},
  \bibinfo {author} {\bibfnamefont {H.}~\bibnamefont {Lamm}}, \bibinfo {author}
  {\bibfnamefont {M.~L.}\ \bibnamefont {Wagman}}, \ and\ \bibinfo {author}
  {\bibfnamefont {N.~C.}\ \bibnamefont {Warrington}},\ }\href {\doibase
  10.1103/PhysRevD.103.094517} {\bibfield  {journal} {\bibinfo  {journal}
  {Phys. Rev. D}\ }\textbf {\bibinfo {volume} {103}},\ \bibinfo {pages}
  {094517} (\bibinfo {year} {2021})},\ \Eprint
  {http://arxiv.org/abs/2101.12668} {arXiv:2101.12668 [hep-lat]} \BibitemShut
  {NoStop}%
\bibitem [{\citenamefont {Beane}\ \emph {et~al.}(2009)\citenamefont {Beane},
  \citenamefont {Detmold}, \citenamefont {Luu}, \citenamefont {Orginos},
  \citenamefont {Parreno}, \citenamefont {Savage}, \citenamefont {Torok},\ and\
  \citenamefont {Walker-Loud}}]{Beane:2009gs}%
  \BibitemOpen
  \bibfield  {author} {\bibinfo {author} {\bibfnamefont {S.~R.}\ \bibnamefont
  {Beane}}, \bibinfo {author} {\bibfnamefont {W.}~\bibnamefont {Detmold}},
  \bibinfo {author} {\bibfnamefont {T.~C.}\ \bibnamefont {Luu}}, \bibinfo
  {author} {\bibfnamefont {K.}~\bibnamefont {Orginos}}, \bibinfo {author}
  {\bibfnamefont {A.}~\bibnamefont {Parreno}}, \bibinfo {author} {\bibfnamefont
  {M.~J.}\ \bibnamefont {Savage}}, \bibinfo {author} {\bibfnamefont
  {A.}~\bibnamefont {Torok}}, \ and\ \bibinfo {author} {\bibfnamefont
  {A.}~\bibnamefont {Walker-Loud}},\ }\href {\doibase
  10.1103/PhysRevD.80.074501} {\bibfield  {journal} {\bibinfo  {journal} {Phys.
  Rev. D}\ }\textbf {\bibinfo {volume} {80}},\ \bibinfo {pages} {074501}
  (\bibinfo {year} {2009})},\ \Eprint {http://arxiv.org/abs/0905.0466}
  {arXiv:0905.0466 [hep-lat]} \BibitemShut {NoStop}%
\bibitem [{\citenamefont {Endres}\ \emph
  {et~al.}(2011{\natexlab{a}})\citenamefont {Endres}, \citenamefont {Kaplan},
  \citenamefont {Lee},\ and\ \citenamefont {Nicholson}}]{Endres:2011jm}%
  \BibitemOpen
  \bibfield  {author} {\bibinfo {author} {\bibfnamefont {M.~G.}\ \bibnamefont
  {Endres}}, \bibinfo {author} {\bibfnamefont {D.~B.}\ \bibnamefont {Kaplan}},
  \bibinfo {author} {\bibfnamefont {J.-W.}\ \bibnamefont {Lee}}, \ and\
  \bibinfo {author} {\bibfnamefont {A.~N.}\ \bibnamefont {Nicholson}},\ }\href
  {\doibase 10.1103/PhysRevLett.107.201601} {\bibfield  {journal} {\bibinfo
  {journal} {Phys. Rev. Lett.}\ }\textbf {\bibinfo {volume} {107}},\ \bibinfo
  {pages} {201601} (\bibinfo {year} {2011}{\natexlab{a}})},\ \Eprint
  {http://arxiv.org/abs/1106.0073} {arXiv:1106.0073 [hep-lat]} \BibitemShut
  {NoStop}%
\bibitem [{\citenamefont {Endres}\ \emph
  {et~al.}(2011{\natexlab{b}})\citenamefont {Endres}, \citenamefont {Kaplan},
  \citenamefont {Lee},\ and\ \citenamefont {Nicholson}}]{Endres:2011mm}%
  \BibitemOpen
  \bibfield  {author} {\bibinfo {author} {\bibfnamefont {M.~G.}\ \bibnamefont
  {Endres}}, \bibinfo {author} {\bibfnamefont {D.~B.}\ \bibnamefont {Kaplan}},
  \bibinfo {author} {\bibfnamefont {J.-W.}\ \bibnamefont {Lee}}, \ and\
  \bibinfo {author} {\bibfnamefont {A.~N.}\ \bibnamefont {Nicholson}},\ }\href
  {\doibase 10.22323/1.139.0017} {\bibfield  {journal} {\bibinfo  {journal}
  {PoS}\ }\textbf {\bibinfo {volume} {LATTICE2011}},\ \bibinfo {pages} {017}
  (\bibinfo {year} {2011}{\natexlab{b}})},\ \Eprint
  {http://arxiv.org/abs/1112.4023} {arXiv:1112.4023 [hep-lat]} \BibitemShut
  {NoStop}%
\bibitem [{\citenamefont {DeGrand}(2012)}]{DeGrand:2012ik}%
  \BibitemOpen
  \bibfield  {author} {\bibinfo {author} {\bibfnamefont {T.}~\bibnamefont
  {DeGrand}},\ }\href {\doibase 10.1103/PhysRevD.86.014512} {\bibfield
  {journal} {\bibinfo  {journal} {Phys. Rev. D}\ }\textbf {\bibinfo {volume}
  {86}},\ \bibinfo {pages} {014512} (\bibinfo {year} {2012})},\ \Eprint
  {http://arxiv.org/abs/1204.4664} {arXiv:1204.4664 [hep-lat]} \BibitemShut
  {NoStop}%
\bibitem [{\citenamefont {Grabowska}\ \emph {et~al.}(2013)\citenamefont
  {Grabowska}, \citenamefont {Kaplan},\ and\ \citenamefont
  {Nicholson}}]{Grabowska:2012ik}%
  \BibitemOpen
  \bibfield  {author} {\bibinfo {author} {\bibfnamefont {D.}~\bibnamefont
  {Grabowska}}, \bibinfo {author} {\bibfnamefont {D.~B.}\ \bibnamefont
  {Kaplan}}, \ and\ \bibinfo {author} {\bibfnamefont {A.~N.}\ \bibnamefont
  {Nicholson}},\ }\href {\doibase 10.1103/PhysRevD.87.014504} {\bibfield
  {journal} {\bibinfo  {journal} {Phys. Rev. D}\ }\textbf {\bibinfo {volume}
  {87}},\ \bibinfo {pages} {014504} (\bibinfo {year} {2013})},\ \Eprint
  {http://arxiv.org/abs/1208.5760} {arXiv:1208.5760 [hep-lat]} \BibitemShut
  {NoStop}%
\bibitem [{\citenamefont {Nicholson}\ \emph {et~al.}(2013)\citenamefont
  {Nicholson}, \citenamefont {Grabowska},\ and\ \citenamefont
  {Kaplan}}]{Nicholson:2012xt}%
  \BibitemOpen
  \bibfield  {author} {\bibinfo {author} {\bibfnamefont {A.~N.}\ \bibnamefont
  {Nicholson}}, \bibinfo {author} {\bibfnamefont {D.}~\bibnamefont
  {Grabowska}}, \ and\ \bibinfo {author} {\bibfnamefont {D.~B.}\ \bibnamefont
  {Kaplan}},\ }\href {\doibase 10.1088/1742-6596/432/1/012032} {\bibfield
  {journal} {\bibinfo  {journal} {J. Phys. Conf. Ser.}\ }\textbf {\bibinfo
  {volume} {432}},\ \bibinfo {pages} {012032} (\bibinfo {year} {2013})},\
  \Eprint {http://arxiv.org/abs/1210.7250} {arXiv:1210.7250 [hep-lat]}
  \BibitemShut {NoStop}%
\bibitem [{\citenamefont {Drut}\ and\ \citenamefont
  {Porter}(2016)}]{Drut:2015uua}%
  \BibitemOpen
  \bibfield  {author} {\bibinfo {author} {\bibfnamefont {J.~E.}\ \bibnamefont
  {Drut}}\ and\ \bibinfo {author} {\bibfnamefont {W.~J.}\ \bibnamefont
  {Porter}},\ }\href {\doibase 10.1103/PhysRevE.93.043301} {\bibfield
  {journal} {\bibinfo  {journal} {Phys. Rev. E}\ }\textbf {\bibinfo {volume}
  {93}},\ \bibinfo {pages} {043301} (\bibinfo {year} {2016})},\ \Eprint
  {http://arxiv.org/abs/1508.04375} {arXiv:1508.04375 [cond-mat.str-el]}
  \BibitemShut {NoStop}%
\bibitem [{\citenamefont {Wagman}\ and\ \citenamefont
  {Savage}(2017{\natexlab{b}})}]{Wagman:2016bam}%
  \BibitemOpen
  \bibfield  {author} {\bibinfo {author} {\bibfnamefont {M.~L.}\ \bibnamefont
  {Wagman}}\ and\ \bibinfo {author} {\bibfnamefont {M.~J.}\ \bibnamefont
  {Savage}},\ }\href {\doibase 10.1103/PhysRevD.96.114508} {\bibfield
  {journal} {\bibinfo  {journal} {Phys. Rev. D}\ }\textbf {\bibinfo {volume}
  {96}},\ \bibinfo {pages} {114508} (\bibinfo {year} {2017}{\natexlab{b}})},\
  \Eprint {http://arxiv.org/abs/1611.07643} {arXiv:1611.07643 [hep-lat]}
  \BibitemShut {NoStop}%
\bibitem [{\citenamefont {Yunus}\ and\ \citenamefont
  {Detmold}(2022)}]{Yunus:2022pto}%
  \BibitemOpen
  \bibfield  {author} {\bibinfo {author} {\bibfnamefont {C.}~\bibnamefont
  {Yunus}}\ and\ \bibinfo {author} {\bibfnamefont {W.}~\bibnamefont
  {Detmold}},\ }\href {\doibase 10.48550/ARXIV.2210.15789} {\enquote {\bibinfo
  {title} {Large-time correlation functions in bosonic lattice field
  theories},}\ }\bibinfo {howpublished} {https://arxiv.org/abs/2210.15789}
  (\bibinfo {year} {2022})\BibitemShut {NoStop}%
\bibitem [{\citenamefont {Stanley}(1968)}]{inception}%
  \BibitemOpen
  \bibfield  {author} {\bibinfo {author} {\bibfnamefont {H.~E.}\ \bibnamefont
  {Stanley}},\ }\href {\doibase 10.1103/PhysRevLett.20.589} {\bibfield
  {journal} {\bibinfo  {journal} {Phys. Rev. Lett.}\ }\textbf {\bibinfo
  {volume} {20}},\ \bibinfo {pages} {589} (\bibinfo {year} {1968})}\BibitemShut
  {NoStop}%
\bibitem [{\citenamefont {Kogut}(1979)}]{kogut}%
  \BibitemOpen
  \bibfield  {author} {\bibinfo {author} {\bibfnamefont {J.~B.}\ \bibnamefont
  {Kogut}},\ }\href {\doibase 10.1103/RevModPhys.51.659} {\bibfield  {journal}
  {\bibinfo  {journal} {Rev. Mod. Phys.}\ }\textbf {\bibinfo {volume} {51}},\
  \bibinfo {pages} {659} (\bibinfo {year} {1979})}\BibitemShut {NoStop}%
\bibitem [{\citenamefont {{de Gennes}}(1972)}]{DEGENNES1972339}%
  \BibitemOpen
  \bibfield  {author} {\bibinfo {author} {\bibfnamefont {P.}~\bibnamefont {{de
  Gennes}}},\ }\href {\doibase https://doi.org/10.1016/0375-9601(72)90149-1}
  {\bibfield  {journal} {\bibinfo  {journal} {Physics Letters A}\ }\textbf
  {\bibinfo {volume} {38}},\ \bibinfo {pages} {339} (\bibinfo {year}
  {1972})}\BibitemShut {NoStop}%
\bibitem [{\citenamefont {van~der Vaart}(2000)}]{Van_der_Vaart2000-fz}%
  \BibitemOpen
  \bibfield  {author} {\bibinfo {author} {\bibfnamefont {A.~W.}\ \bibnamefont
  {van~der Vaart}},\ }\href@noop {} {\emph {\bibinfo {title} {Cambridge series
  in statistical and probabilistic mathematics: Asymptotic statistics series
  number 3}}}\ (\bibinfo  {publisher} {Cambridge University Press},\ \bibinfo
  {address} {Cambridge, England},\ \bibinfo {year} {2000})\BibitemShut
  {NoStop}%
\bibitem [{\citenamefont {Romero}(2020)}]{agimenezromero}%
  \BibitemOpen
  \bibfield  {author} {\bibinfo {author} {\bibfnamefont {A.}~\bibnamefont
  {Romero}},\ }\href@noop {} {\enquote {\bibinfo {title} {Phi-4-model},}\
  }\bibinfo {howpublished}
  {\url{https://github.com/Physics-Simulations/Phi-4-Model}} (\bibinfo {year}
  {2020})\BibitemShut {NoStop}%
\bibitem [{\citenamefont {Keener}(2010)}]{Keener2010}%
  \BibitemOpen
  \bibfield  {author} {\bibinfo {author} {\bibfnamefont {R.~W.}\ \bibnamefont
  {Keener}},\ }\href {\doibase 10.1007/978-0-387-93839-4} {\emph {\bibinfo
  {title} {Theoretical Statistics}}}\ (\bibinfo  {publisher} {Springer New
  York},\ \bibinfo {year} {2010})\BibitemShut {NoStop}%
\bibitem [{\citenamefont {L\"uscher}(2018)}]{Luscher:2017cjh}%
  \BibitemOpen
  \bibfield  {author} {\bibinfo {author} {\bibfnamefont {M.}~\bibnamefont
  {L\"uscher}},\ }\href {\doibase 10.1051/epjconf/201817501002} {\bibfield
  {journal} {\bibinfo  {journal} {EPJ Web Conf.}\ }\textbf {\bibinfo {volume}
  {175}},\ \bibinfo {pages} {01002} (\bibinfo {year} {2018})},\ \Eprint
  {http://arxiv.org/abs/1707.09758} {arXiv:1707.09758 [hep-lat]} \BibitemShut
  {NoStop}%
\end{thebibliography}%

\end{document}